%% file: 6182.tex
\def\to{\hbox{$\,$--$\,$}}
\def\muspc{\hskip 0.15 em}

\def\mag{\hbox{$\;.\!\!\!^m$}}
\documentstyle{l-aac}
\input psfig
\begin{document}

\thesaurus{06(3.13.2, 08.08.1, 08.19.1, 10.19.1)}

\title{Revised ages for stars in the solar
neighbourhood\thanks{Tables 3\to8 are available in electronic
form at the CDS via anonymous {\tt ftp} to
{\tt cdsarc.u-strasbg.fr (130.79.128.5)}
or {\tt WWW} at
{\tt URL http://cdsweb.u-strasbg.fr/Abstract.html}
}}

\author{Y.K. Ng$\rm^{1}$, G. Bertelli$\rm^{2,3}$}

\institute{
$\rm ^{1}$ Padova Astronomical Observatory,
Vicolo dell'Osservatorio 5, I-35122 Padua, Italy \\
$\rm ^{2}$ Department of Astronomy,
Vicolo dell'Osservatorio 5, I-35122 Padua, Italy \\
$\rm ^{3}$ National Council of Research, CNR -- GNA, Rome, Italy \\
$\rm ^{\null\ }$ E-mail:\quad {\tt Bertelli, Yuen
\char64\ astrpd.pd.astro.it}}

\date{Received 23 January 1997 / Accepted 1 July 1997}

\maketitle

\markboth{Y.K. Ng \& G. Bertelli:\ Revised ages for stars
in the solar neighbourhood}
{}

\begin{abstract}
New ages are computed for the stars from the Edvardsson et~al.\ (1993)
data set. The revised values are systematically larger
toward older ages ($t\!>\!4$~Gyr), while they are slightly lower
for $t\!<\!4$~Gyr.
A similar, but considerably smaller trend is present when
the ages are computed with the distances 
based on {\it Hipparcos}\/ parallaxes.
The resulting age-metallicity relation has a small, but distinct slope
of \mbox{$\sim$\muspc0.07~dex/Gyr}.
\keywords{methods: data analysis -- Stars: statistics, HR-Diagram
-- Galaxy: solar neighbourhood}
\end{abstract}

\section{Introduction}
The Edvardsson et~al.\ (1993, hereafter Edv93ea) sample of 
F- \& G-stars show a slowly, increasing gradient with decreasing age
in the age-metallicity relation (AMR, see Edv93ea Fig.~14a).
There is mainly a large scatter in metallicity at a particular age
or vice versa.
This scatter could be real or it resides either partly in
the metallicity or in the age determined for each star.
\par
We focus the present analysis on the age.
A bias in the ages could originate from the isochrones or
the method adopted for the age determination.
Edv93ea used the Vandenberg (1985, hereafter Vdb85) isochrones
to determine the age in the ($\log T_{\rm eff},\Delta M_V$) diagram.
We used the isochrones from Bertelli et~al.\ (1994, hereafter Bert94ea)
to determine the age in the ($\log T_{\rm eff},M_V$) diagram,
see Sect.~2.2.2 for details.
Moreover, the new distances from {\it Hipparcos}\/ 
parallaxes (ESA 1997) further justify a re-examination of the stellar ages
of the Edv93ea sample.    
\par
The aim of this paper is to determine new ages for
the stars from the Edv93ea sample. Section~2 gives a brief
description of the data set, an outline of the method used
to determine the ages of the stars and the results.
In Sect.~3 we present the
newly obtained AMR and discuss the limitations of the analysis.

\section{Analysis}
\subsection{Data}
The sample comprises mainly a
volume limited \mbox{($d\!<$\muspc50~pc)} set of F- \& G-type stars
for which the distances are determined (Edv93ea -- photometric 
distances; ESA 1997 -- trigonometric parallaxes) and
the effective temperatures \& metallicities are determined
from spectra (see Edv93ea for details).
The \mbox{V-band} photometry of these stars are taken from
Gr{\o}nbech{\muspc\&\muspc}Olsen (1976), Olsen (1983, 1993),
Perry, Olsen{\muspc\&\muspc}Crawford (1987), 
Schuster{\muspc\&\muspc}Nissen (1988),
and Hoffleit{\muspc\&\muspc}Warren (1991).

\subsection{Method}
\subsubsection{The isochrones}
We used the Bert94ea isochrones for the determination of
stellar ages from the Edv93ea sample.
The initial chemical composition of the isochrones
parameterized with ($Y,Z$)
obeys the relation $\Delta Y/\Delta Z=2.5$ (Pagel 1989).
The grid of isochrones is obtained from stellar models
computed with the radiative opacities from OPAL (Iglesias et~al.\ 1992,
Rodgers{\muspc\&\muspc}Iglesias 1992). They have ($Y,Z$): (0.23,0.0004),
(0.24,0.004), (0.25,0.008), (0.28,0.02), and (0.352,0.05).
One isochrone grid with $(Y,Z)=(0.23,0.001)$ is based on models calculated
with the radiative opacities from Huebner et~al.\ (1977).
This complete set of isochrones of fixed metallicity is
interpolated to obtain isochrones with intermediate metallicities.
The metallicity range, smoothly covered by the isochrones,
thus spans from \mbox{$Z$\muspc=\muspc0.0004\to0.05}.
\par
\begin{table}[t]
\caption{Sensitivity of the age determination through
variation of the input parameters (effective temperature,
metallicity, distance).
Bias corrections (Lutz\mbox{\muspc\&\muspc}Kelker 1973, see Sect.~2.2.5 for
additional details) were not applied.
For each parameter with a good fit to the data
the age difference $\delta$ is calculated with
respect to the mean age obtained for the star,
i.e. $\delta=\rm log(age_*)-\overline{log(age_*)}$.
The table gives the mean age difference $\overline\delta$
for each parameter together with its variance $\sigma$.
The first set demonstrates the sensitivity of the parameters
with the uncertainties as discussed in Sect.~2.2.3.
In the second and third set of calculations
we assume respectively a 10\% and 5\% uncertainty in the distance.
In the last set we assume also a
smaller uncertainty for $T_{\rm eff}$ and [Fe/H] for a 10\%
uncertainty in the distance
}
\begin{center}
\begin{tabular}{|l|rrrr|}
\hline
variation & $\overline{\delta_{MS}}$ & $\sigma_{MS}$
& $\overline{\delta_{SGB}}$ & $\sigma_{SGB}$ \\
\hline
none                            &  0.03 & 0.19 &  0.02 & 0.09 \\
$T_{\rm eff}-50$~K              &  0.10 & 0.12 &  0.05 & 0.10 \\
$T_{\rm eff}+100$~K             &--0.11 & 0.32 &--0.05 & 0.08 \\
$\lbrack{\rm Fe/H}\rbrack-0.07$ &  0.10 & 0.10 &  0.06 & 0.08 \\
$\lbrack{\rm Fe/H}\rbrack+0.07$ &--0.01 & 0.09 &--0.01 & 0.08 \\
$d_*-15\%$                      &--0.17 & 0.40 &--0.04 & 0.46 \\
$d_*+15\%$                      &  0.07 & 0.17 &--0.04 & 0.10 \\
\hline
none                            &  0.04 & 0.21 &  0.01 & 0.04 \\
$T_{\rm eff}-50$~K              &  0.11 & 0.17 &  0.04 & 0.05 \\
$T_{\rm eff}+100$~K             &--0.10 & 0.30 &--0.06 & 0.06 \\
$\lbrack{\rm Fe/H}\rbrack-0.07$ &  0.11 & 0.15 &  0.05 & 0.05 \\
$\lbrack{\rm Fe/H}\rbrack+0.07$ &--0.04 & 0.12 &--0.03 & 0.04 \\
$d_*-10\%$                      &--0.21 & 0.68 &  0.02 & 0.10 \\
$d_*+10\%$                      &  0.08 & 0.22 &--0.02 & 0.06 \\
\hline
none                            &  0.02 & 0.19 &  0.01 & 0.04 \\
$T_{\rm eff}-50$~K              &  0.09 & 0.15 &  0.04 & 0.04 \\
$T_{\rm eff}+100$~K             &--0.13 & 0.31 &--0.07 & 0.06 \\
$\lbrack{\rm Fe/H}\rbrack-0.07$ &  0.09 & 0.13 &  0.05 & 0.04 \\
$\lbrack{\rm Fe/H}\rbrack+0.07$ &--0.02 & 0.07 &--0.03 & 0.04 \\
$d_*-5\%$                       &--0.08 & 0.40 &  0.01 & 0.05 \\
$d_*+5\%$                       &  0.05 & 0.15 &  0.00 & 0.05 \\
\hline
none                            &  0.03 & 0.16 &  0.01 & 0.03 \\
$T_{\rm eff}-25$~K              &  0.08 & 0.16 &  0.02 & 0.04 \\
$T_{\rm eff}+50$~K              &--0.03 & 0.13 &--0.02 & 0.04 \\
$\lbrack{\rm Fe/H}\rbrack-0.04$ &  0.08 & 0.14 &  0.03 & 0.04 \\
$\lbrack{\rm Fe/H}\rbrack+0.04$ &  0.00 & 0.22 &--0.02 & 0.03 \\
$d_*-10\%$                      &--0.22 & 0.69 &  0.01 & 0.10 \\
$d_*+10\%$                      &  0.08 & 0.23 &--0.02 & 0.06 \\
\hline
\end{tabular}
\end{center}
\end{table}
%
%
\begin{table}[t]
\caption{Similar to Table~1, but now with 
the distances derived from the  
{\it Hipparcos}\/ parallaxes (ESA 1997). 
}
\begin{center}
\begin{tabular}{|l|rrrr|}
\hline
variation & $\overline{\delta_{MS}}$ & $\sigma_{MS}$
& $\overline{\delta_{SGB}}$ & $\sigma_{SGB}$ \\
\hline
none                            &  0.01 & 0.17 &  0.01 & 0.03 \\
$T_{\rm eff}-50$~K              &  0.13 & 0.27 &  0.04 & 0.04 \\
$T_{\rm eff}+100$~K             &--0.21 & 0.69 &--0.07 & 0.06 \\
$\lbrack{\rm Fe/H}\rbrack-0.07$ &  0.12 & 0.22 &  0.05 & 0.04 \\
$\lbrack{\rm Fe/H}\rbrack+0.07$ &--0.05 & 0.16 &--0.04 & 0.04 \\
$\pi+\sigma_\pi$                &--0.01 & 0.18 &  0.01 & 0.04 \\
$\pi-\sigma_\pi$                &  0.01 & 0.28 &  0.00 & 0.04 \\
\hline
\end{tabular}
\end{center}
\end{table}
This enables us to generate isochrones with
metallicities comparable to those used by Edv93ea.
There are however marked differences
between the Vdb85 and Bert94ea
isochrones. They are respectively:
\begin{list}{$-$}{\topsep=0pt\parsep=0pt}
\item the helium fraction, fixed $Y=0.25$
versus variable through $\Delta Y/\Delta Z=2.5$;
\item the radiative opacities,
LAOL (Huebner et~al.\ 1977) versus
OPAL (Iglesias et~al.\ 1992, Rodgers{\muspc\&\muspc}Iglesias 1992);
\item initial metal abundance mix of elements heavier than helium,
Vandenberg (1983) versus Grevesse (1991);
\item convective overshoot, none versus included.
\end{list}
In addition, different tables have been used by Vdb85 and Bert94ea
to convert the effective temperature and luminosity into
magnitude and colours. The Vdb85 isochrones
were computed for metal mass fractions of
$Z$\muspc=\muspc0.0169, 0.01, 0.006, 0.003, and 0.0017.
Note that for the conversion of [Fe/H] or [Me/H] to $Z$
the reference solar metallicity is
\mbox{$Z_\odot$(Vdb85)\muspc=\muspc0.0169} and
\mbox{$Z_\odot$(Bert94ea)\muspc=\muspc0.020}.
In addition the Vdb85 isochrones ought to be shifted
by \mbox{$\delta\log T_{\rm eff}$\muspc=\muspc--0.009}
to satisfy the solar constraint (see Edv93ea p119 for details).
Figure~1 shows a set of \mbox{[Fe/H]\muspc=\muspc0.0} isochrones
from Vdb85 and Bert94ea.
Note the different appearance of the 10~Gyr isochrone from Bert94ea
with respect to isochrones of younger age
due to the presence of convective overshoot in the latter.
\par

\subsubsection{Fitting ($\log T_{\rm eff},M_V$)}
For each star in the sample the metallicity is known.
Each star can be located in a Hertzsprung-Russell diagram (HRD)
through its effective temperature and absolute visual magnitude.
We used the interpolated isochrones from Bert94ea
to determine the age of the star, where the stellar mass
specifies the position along the selected isochrone.
Powell's method (Press et~al.\ 1986) is used
to obtain the stellar age and mass. We
minimize the difference
in the ($\log T_{\rm eff},M_V$) diagram
between the observed star and the
interpolated isochrones from Bert94ea
(see for example Fig.~1).
For this purpose we defined
a weighted, reduced chi-squared
$$
\chi^2_{\rm r}={\sqrt{
\Big(M_{V,o}-M_{V,i}\Big)^2
+ 16\Big(\log T_{{\rm eff},o}-\log T_{{\rm eff},i}\Big)^2
}\over\sigma(M_{V,o})} \;.
$$
The subscripts {\tt o} and {\tt i} refer respectively to
the observed and the synthetic, isochrone quantities.
We used the Bert94ea relation --- $\log Z/Z_\odot=0.977{\rm [Fe/H]}$ ---
to convert [Fe/H] to $Z$.
\par
\begin{figure}
\centerline{\vbox{\psfig{file=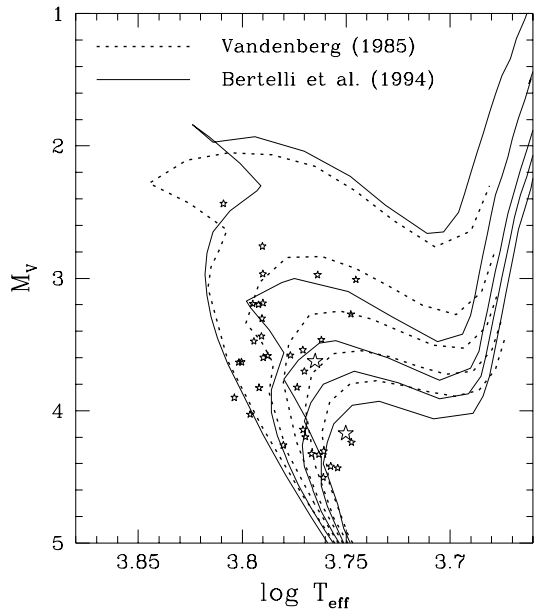,height=8.0cm,width=7.05cm}
}}
\caption{Comparison for [Fe/H]=0.0
between the Vdb85, LAOL based isochrones
with the OPAL based isochrones from Bert94ea.
The isochrones ages are 2, 4, 6, 8 and 10 Gyr for the LAOL
based set and 2, 4, 6.3, 8 and 10 Gyr for the OPAL based set.
The Vdb85 isochrones are shifted by
\mbox{$\delta\log T_{\rm eff}$\muspc=\muspc--0.009}
to satisfy the solar constraint (see Edv93ea for details).
In addition, the stars (Table~4) with metallicities
$-0.05\!<\! {\rm [Fe/H]}\!<\!0.05$ have been inserted. The symbols are
the same as those in the following figures}
\label{isoc}
\end{figure}
Note, that our fitting method is not substantially
different from the determination of the age in the
($\log T_{\rm eff},\Delta M_V$) diagram used by Edv93ea. Since
$\Delta M_{V,o}$\muspc=\muspc$M_{V,o,zams}-M_{V,o}$
and
$\Delta M_{V,i}$\muspc=\muspc$M_{V,i,zams}-M_{V,i}$,
then minimizing the difference between the observed and
the isochrone magnitude implies:
$\delta M_V=\Delta M_{V,o}-\Delta M_{V,i}$.
With $M_{V,o,zams}=M_{V,i,zams}$ this equals
to the quantity minimized in the reduced chi-squared defined above.
If on the other hand $M_{V,o,zams}\ne M_{V,i,zams}$
then an unnecessary bias would have been added in the results.
In that case minimization in the ($\log T_{\rm eff},M_V$) plane
should be preferred over
minimization in the ($\log T_{\rm eff},\Delta M_V$) plane.
\par

\subsubsection{Uncertainties in $T_{\rm eff}$, [Fe/H] \& distance}
The uncertainties in the effective temperature, metallicity
and the distance will give different results
for the fitted mass and age along the isochrone. For each star
we repeated the calculations with slightly different
values for the effective temperature
(--50/+100~K; see Edv93ea),
the metallicity ($\pm0.07$ in [Fe/H]\/; adopted) and
the distance ($\pm15\%$ for distances given by Edv93ea; 
in Tables~7\muspc\&\muspc8 the uncertainties are considered separately for each
star with an {\it Hipparcos}\/ parallax).
In total 7 different values for the mass and the age
of each star are obtained.
A consequence of the automatic fitting procedure is that it
always provides an answer, which sometimes is not a good
fit to the actual data. The latter were not considered in the
calculation of the unweighted mean
and the uncertainty in the age and mass of each star in the sample.
\par
\begin{figure}
\centerline{\vbox{\psfig{file=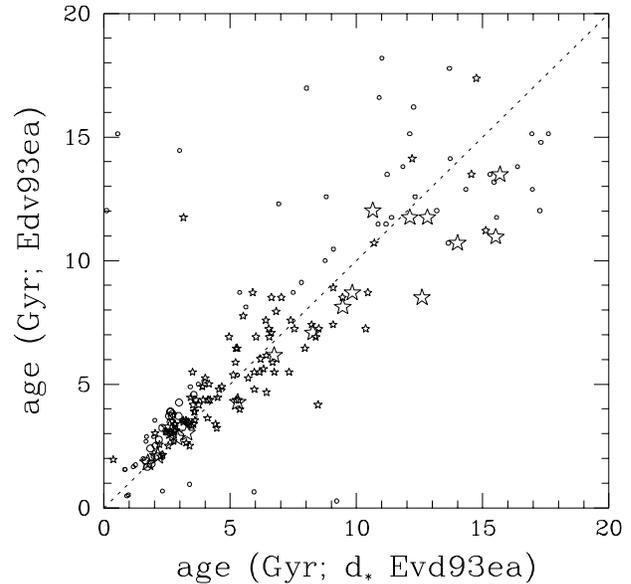,height=7.6cm,width=8.0cm}
}}
\caption{A comparison between the ages from Tables~3\muspc\&\muspc4 with
those determined by Edv93ea,
due to different isochrone sets.
The open circles denote the MS stars from Table~3 and 
the asterisks denote the SGB stars from Table~4.
The big symbols are for stars with at least 6 out of the
maximum 7 good age obtained from the analysis
(see Sect.~2.2.3 for details) and for which
the uncertainty in $\rm\overline{log(age)}$ is
smaller than 0.05. The small symbols are used in all remaining cases
and refer in general to stars with less reliable ages
}
\label{comp_ngedv}
\end{figure}
When various measures for a particular
star of the V-band photometry were found
in different databases the complete procedure as described
above was applied for each photometric measurement.
Then an average was computed, weighted over the
photometric errors given by the various authors
or adopted by us (0\mag01 for the photometry
from the bright star catalogue and 0\mag05 if the photometric errors
were not specified at all).
\par

\subsubsection{Isochrone population}
The method thus far considers only the formal solution, automatically
generated by the procedure outlined above.
However, ambiguities in the age determinations are not taken into
account when the stars are located near the end of the core-H
burning phase. In the vicinity of the termination point (i.e. the
hook feature)
of young isochrones, it is possible that the star is either
at the end of the main sequence phase or beyond the termination point,
each implying a different value for the age of the star.
\hfill\break
In the mass range from 1.0\to1.7~M$_\odot$
the evolution along the SGB (Sub-Giant Branch) is
roughly 4\to40 times faster
than the evolution from the ZAMS (Zero Age Main Sequence)
up to the termination point.
Evolved stars must therefore be inspected individually. If the
resulting age results in a position on the isochrone beyond the
termination point, then an alternative age corresponding to a point
below the termination point is assigned to the star
(see last column Table~4).
\par
\begin{figure}
\centerline{\vbox{\psfig{file=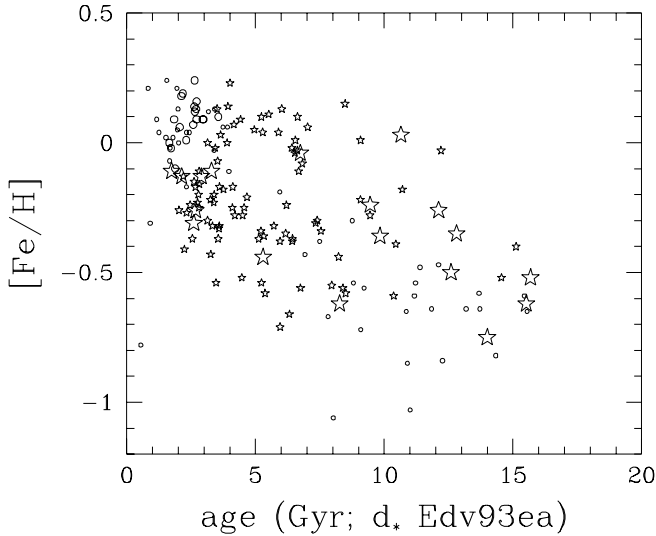,height=6.91cm,width=8.5cm}
}}
\caption{The AMR for all the stars in
Tables 3\&4. See Fig.~2 for the symbol description,
but the small symbols are used here for the remaining cases
with $n\!>\!2$ and an uncertainty in $\rm\overline{log(age)}\!<\!0.3$}
\label{amr_ng_nolk}
\end{figure}
\begin{figure}
\centerline{\vbox{\psfig{file=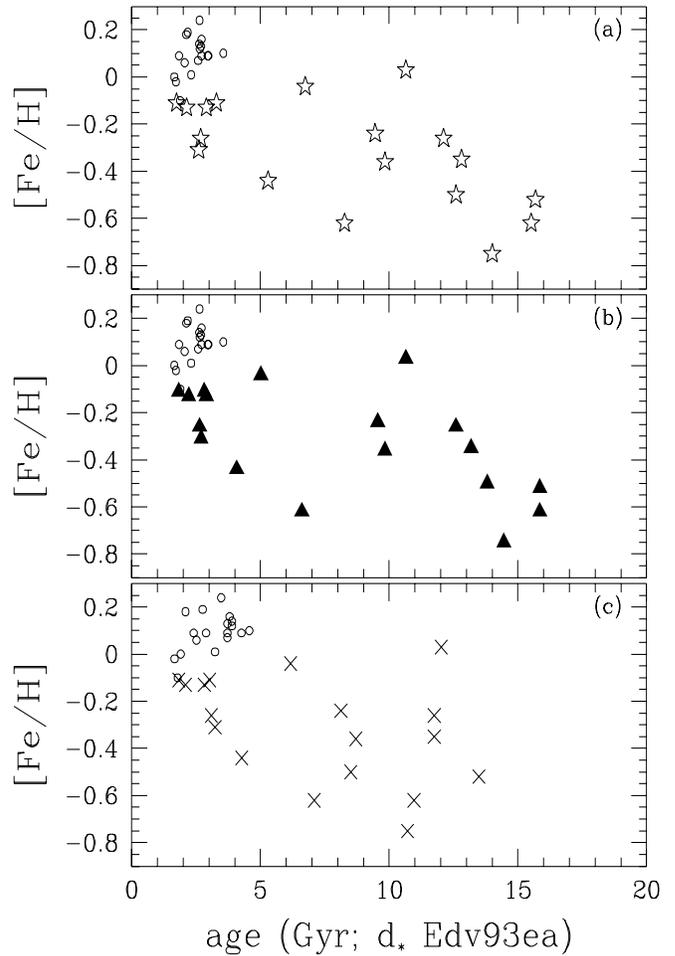,height=12.6cm,width=8.5cm}
}}
\caption{({\bf a}) The AMR for the stars in
Tables~3\&4 with a dispersion in
$\rm\overline{log(age)}\!<\!0.05$
(see Fig.~2 for the symbol description),
({\bf b}) the AMR for the same stars, but now with the
`visually' assigned ages for the SGB stars (filled triangles,
last column Table~4),
and ({\bf c}) the AMR for the same stars, but now with the ages from
Edv93ea.}
\label{amr_select}
\end{figure}
%
%

\subsubsection{The Lutz-Kelker correction}
Lutz\mbox{\muspc\&\muspc}Kelker (1973) demonstrated that a bias
is present among stars with their absolute magnitudes obtained 
from trigonometric parallax. The bias is not confined
to volume limited samples as discussed by 
Trumpler\mbox{\muspc\&\muspc}Weaver (1953) and it
depends only on the ratio ($\sigma_\pi/\pi$), i.e. the ratio of the 
standard error of the parallax over the parallax. 
Due to observational
errors in the parallax one expects to find 
statistically more stars from a larger distance scattering into a smaller
volume than vice versa. The consequence of this geometrical effect
is that the stars in a sample are on average brighter and 
younger than they appear to be.
\par
Thus far we have considered the formal uncertainties in the
method to determine the age of the stars. Edv93ea compared 
their photometrically derived distances with those obtained from
ground based parallaxes.
They found an excellent agreement and estimated that the uncertainty
in the distances is about 15\%. 
Their {\it\`a posteriori}\/ information about the compatibility with 
parallax data 
implies that we ought to apply a Lutz-Kelker correction
in our analysis. This correction is in the generalized 
case only invoked by the relative uncertainties in the distance scale. 
Note however, that Edv93ea (p121 and references cited therein) do not have 
to apply a Lutz-Kelker correction, because the distances 
that we adopted from their paper were neither required 
or used in their $\Delta M_V$-method.  
\par
The Lutz-Kelker correction is applied 
to the absolute magnitudes determined from 
distances for the stars from Edv93ea.
We expect, within the volume considered, a homogeneous 
distribution for the F- \& G-type stars 
and we used Hanson's equation 31 (Hanson 1979) with
index $n\!=\!4$ to compute the value of the correction. 
The stars become about 0\mag3 brighter
with $\sigma_\pi/\pi\!\simeq\!0.15$.
The correction leads to negligible differences 
in the results with distances from the {\it Hipparcos}\/ parallax,
because for the majority of the stars $\sigma_\pi/\pi\!<\!0.03$.
\par

\subsection{Results}
The results are presented in three stages. First, we
show the differences due to a change
of isochrones: Vdb85 versus Bert94ea.
However, no corrections were made 
for a bias in the absolute magnitudes.
This is done in the second stage of the analysis. 
In the third and final stage we based the analysis on the 
{\it Hipparcos}\/ parallax. 
\par
\begin{figure}[t]
\centerline{\vbox{\psfig{file=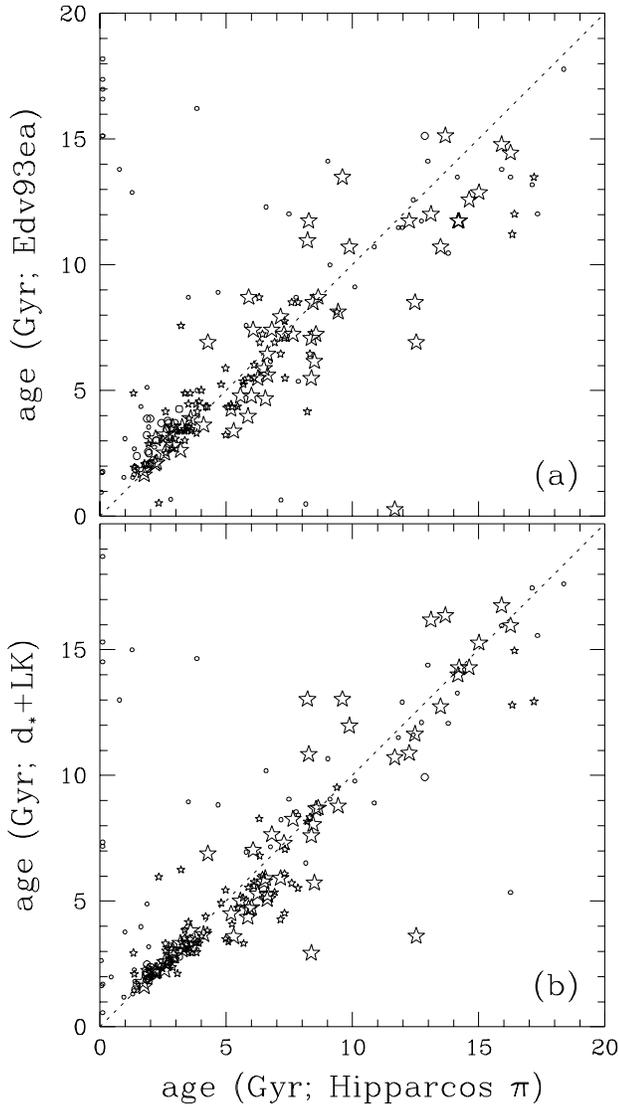,height=14.7cm,width=8.0cm}
}}
\caption{A comparison of the ages computed with 
the distances of the stars based on {\it Hipparcos}\/ 
parallaxes (Tables~7\mbox{\muspc\&\muspc}8). 
Panel {\bf (a)} shows the comparison 
with the ages from Edv93ea. The differences are due 
to the isochrones and the fitting method.
In panel {\bf (b)} we make a comparison with the ages from
Tables~5\mbox{\muspc\&\muspc}6.
The differences are only due the distances adopted for the stars 
(photometric distances from Edv93ea with the Lutz-Kelker correction
versus distances from trigonometric parallax).
Symbols are the same as described in Fig.~2
}
\label{compare_hipparcos}
\end{figure}
\begin{figure}[t]
\centerline{\vbox{\psfig{file=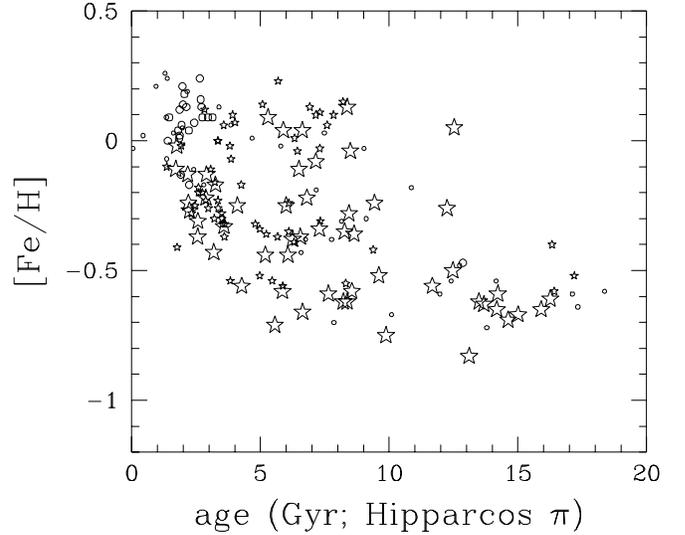,height=6.91cm,width=8.5cm}
}}
\caption{The AMR similar to Fig.~3, but now for all the stars in
Tables~7\mbox{\muspc\&\muspc}8 with their distances based on 
{\it Hipparcos}\/ parallaxes (ESA 1997). 
See Fig.~2 for the symbol description.
Small symbols are used here for the remaining cases
with $n\!>\!2$ and an uncertainty in $\rm\overline{log(age)}\!<\!0.3$}
\end{figure}
 \par
Table~1 shows the sensitivity of the age through
individual variation of each of the input parameters
in comparison with the mean age obtained for each star.
The first set of calculations show the sensitivity of the
results for the uncertainties, as described
in Sect.~2.2.3, adopted for the input parameters.
In the second \& third set of calculations a smaller
uncertainty in the distance was adopted and in the fourth
set we further assumed smaller uncertainties in the
effective temperature and metallicity.
\hfill\break
The average results, 
from the procedure outlined in preceding sections,
are given in the Tables~3\mbox{\muspc\&\muspc}4. 
They list respectively
the stars which are on or very near to the MS (Main Sequence)
and those which are on the SGB.
The Lutz-Kelker correction was not applied in order 
to discuss the effects from different
sets of isochrones. 
Tables~5{\muspc\&\muspc}6 contain the ages when the 
correction is applied.
\hfill\break
Table 2 shows, similar to Table~1, the sensitivity of the age
through variation of the individual parameters, but now
based on {\it Hipparcos}\/ distances instead of the distances
given by Edv93ea. The
average results for each star (MS and SGB) are given in the 
Tables~7\mbox{\muspc\&\muspc}8.
\par
In the following set of figures we first compare the difference 
due to a different choice of isochrones, i.e. Figs.~2\to4. 
Figs.~5ab\mbox{\muspc\&\muspc}6 show the relations
when the trigonometric parallaxes from {\it Hipparcos}\/ are adopted.
\hfill\break
In Fig.~2 we compare the ages for the stars computed with the
Bert94ea isochrones with the ages given by Edv93ea. The comparison 
shows that towards older age
the revised values are
systematically larger than Edv93ea, see Sect.~3.1
for a detailed discussion about this difference.
Note that we refer to the stars with reliable ages, i.e.
the big symbols, and discarded the old MS stars (small open
circles) from the comparison.
\hfill\break
Figure~3 shows the age-metallicity for all the stars listed
in Tables~3\mbox{\muspc\&\muspc}4. Figure~4a shows the AMR
for the SGB stars with an average age obtained from
at least six out of the maximum of seven
good age determinations with an uncertainty in
\mbox{log(age)\muspc$<$\muspc0.05}. Figures~4b\mbox{\muspc\&\muspc}4c 
show for comparison the AMR
for the same stars, but now with respectively
the `visually'
assigned ages for the SGB stars (see also Sect.~2.2.4)
and the ages from Edv94ea.
\hfill\break
In Figs.~5a\mbox{\muspc\&\muspc}5b we compare the 
ages (Tables~7\mbox{\muspc\&\muspc}8), based
on absolute magnitudes obtained from {\it Hipparcos}\/
parallaxes,
with those obtained by Edv93ea and the ages given 
in Tables~5\mbox{\muspc\&\muspc}6. 
Figure~6 is similar to Fig.~3,
but now the ages are based on the distances from the {\it Hipparcos}\/ parallax
instead of the distances given by Edv93ea.
\hfill\break
Figure~7a displays the relation obtained
from distances derived from the {\it Hipparcos}\/ parallax, while
Fig.~7b shows for comparison the same stars, but now
with the ages from Edv93ea.
\par

\section{Discussion}

\subsection{Revised ages}
\subsubsection{Isochrones}
In Fig.~2 a comparison is made between the ages obtained by Edv93ea
and those determined here with the Bert94ea isochrones.
The general trend is that the revised ages get
systematically larger towards older ages.
On the other hand, the revised ages of a few MS stars
are found to be considerably younger, while a few other young MS
stars are now considerably older.
\hfill\break
The general trend should mainly originate from the
differences between the Vdb85 and Bert94ea isochrones.
Differences in the opacities, initial abundance mix,
and the different conversion tables have barely influence
on the age obtained.
\hfill\break
$\bullet$\quad
Convective overshoot is only present in
stars with $M\!\ge\!1\;M_\odot$ or \mbox{$t\!\la\!10$~Gyr}
(see Chiosi et~al.\ 1992,
Bert94ea, and references cited in those papers).
The effect of convective overshoot at the MS is that
the stars have larger cores. This results in a higher
luminosity and a prolonged age. It should lead to
systematically older ages for the stars
with respect to ages determined
from the Vdb85 isochrones.
\hfill\break
$\bullet$\quad
The differences in the helium mass fraction has the following
effect on the age determination:
\begin{list}{$-$}{\topsep=0pt\parsep=0pt}
\item for solar metallicity isochrones the helium mass fraction $Y$ is lower
in the Vdb85 isochrones; the effects of difference in $Y$ was checked with
the isochrones present in our data base with
\mbox{($Z$\muspc=\muspc0.05,$Y$\muspc=\muspc0.352)} and
\mbox{($Z$\muspc=\muspc0.05,$Y$\muspc=\muspc0.4}, unpublished);
a lower $Y$ leads to a lower $T_{\rm eff}$
due to an increased opacity from a higher
hydrogen mass fraction $X$, while the turn-off luminosity remains
comparable; as a result the MS turn-off ages
determined with isochrones with a lower $Y$ will be younger;
from this argument one expects older ages from
the Bert94ea isochrones;
\item for metal-poor isochrones $Y$ is higher in the Vdb85 isochrones;
this leads to the reverse effect, i.e. younger ages
from the Bert94ea isochrones.
\end{list}
The general trends outlined above are 
at young age not in agreement with Fig.~2, because we expect older ages
from the Bert94ea isochrones. Instead we find a comparable or even a slightly
younger age. A similar discrepancy is found for old stars 
with relatively low metallicities.
We further note, that the
Vdb85 isochrones with \mbox{[Fe/H]\muspc=\muspc0.0}
were not as expected from the difference in $Y$
located at lower, but higher effective temperature.
In addition, the shift of 
\mbox{$\delta\log T_{\rm eff}$\muspc=\muspc$-\!0.009$}
applied by Edv93ea to the Vdb85 isochrones
makes it even more difficult
to determine the origin of the differences between the Edv93ea
and our revised ages.
\hfill\break
There are too many differences (e.g. opacities, convective overshoot,
normalization, conversion tables ...) between the Bert94ea and the 
Vdb85 isochrones
that it is not only hard to compare them, but also to understand
which of the differences or a combination of them
is/are the dominant factor(s).
All the factors mentioned above hamper the analysis
on the origin of the differences between the Edv93ea and the revised ages.
Although we are able to outline general differences
between the re-normalized Vdb85 and Bert94ea isochrones,
we fail to prove with plain arguments
the exact differences between two isochrone sets
and hence the differences between the ages
obtained by Edv93ea and this work.
To understand the difference between these isochrones requires 
a deep and detailed analysis of isochrones
with the same $Y,Z$. Unfortunately they are not available to us,
while their computation is beyond the scope of this paper.


\subsubsection{{\sl Hipparcos}\/ parallaxes}
Figure~5a shows the effects due to differences in the 
isochrones and the fitting method.
The reliable ages of the stars based on the {\it Hipparcos}\/ 
parallax (the big symbols) get \mbox{$\sim$\muspc2~Gyr} older for stars
in the age spanning \mbox{9\to16~Gyr},
\mbox{$\sim$\muspc1~Gyr} older for stars
in the age spanning \mbox{4\to9~Gyr}, 
and a negligible difference for stars younger than
4~Gyr. 
Figure 5b displays the effects due to differences in
the distance: photometric distances from Edv93ea and the 
derived absolute magnitudes
corrected for the Lutz-Kelker bias
versus the distances derived from the
{\it Hipparcos}\/ parallax.
It indicates that, besides some increased scatter,
the ages based on the {\it Hipparcos}\/
parallax are slightly younger for $t\!>\!8$~Gyr.
\par
\begin{figure}
\centerline{\vbox{\psfig{file=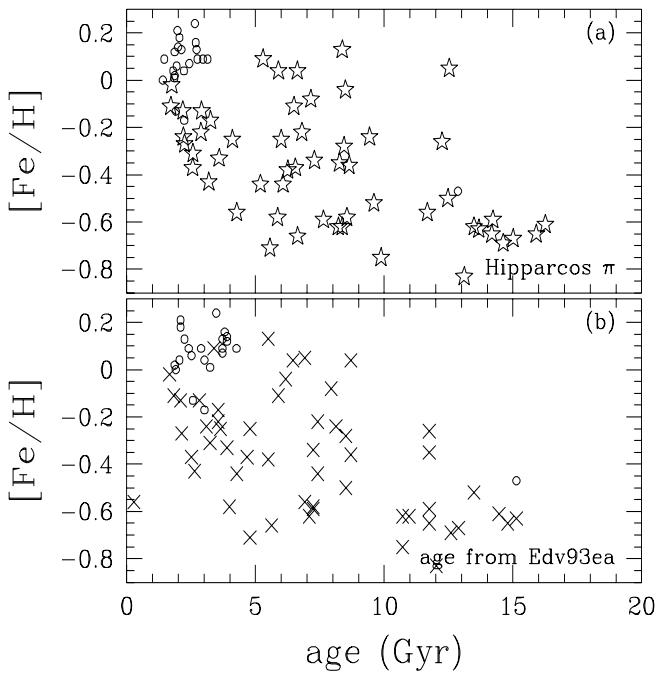,height=8.8cm,width=8.5cm}
}}
\caption{{\bf (a)} The AMR for the stars in
Tables~7\mbox{\muspc\&\muspc}8 with a dispersion in
$\rm\overline{log(age)}\!<\!0.05$
(see Fig.~2 for the symbol description)
and {\bf (b)} the AMR for the same stars, 
but now with the ages from Edv93ea.
}
\label{amr_selectb}
\end{figure}

\subsection{Sensitivity of the results}
Table~1 shows the results from four sets of calculations
to check the sensitivity of the results caused by
uncertainties in the input parameters.
The mean age difference $\overline\delta$ 
gives the shift of the age due to
variation of each of the input parameters. 
The variance $\sigma$ gives an indication
about the scatter in the input parameter varied.
\par
All sets of calculations indicate that both $\overline\delta$
and $\sigma$ are significantly larger for the MS stars.
This shows that the stars which contribute mostly to a large
age spread are automatically identified as MS stars. 
The analysis further shows that the 15\% uncertainty in the distance 
of the SGB stars is the main cause of their large age spread. An uncertainty
of 5\% is required to obtain variances comparable to those 
obtained for the other input parameters. 
A better determination of $T_{\rm eff}$ and/or [Fe/H]
does not lead to a significant improvement as
long as the distances are not accurate down to a 5\% level.
\hfill\break
Table~2 shows that a slightly larger improvement is obtained 
when the {\it Hipparcos}\/ parallaxes are used.
A reduction of the uncertainty in the ages will not be obtained  
from even more accurate distances, but from a better definition
of the effective temperatures for the stars in the sample. 
Figure~7a further demonstrates that the number
of stars with reliable ages has increased considerable with
respect to the stars displayed in Fig.~4a.
\par

\subsection{MS stars}
Figures 3 and 6 show the age-metallicity of all the
stars in the sample analysed. 
The uncertainties in the ages of old MS stars are quite large.
Near the MS the evolution is quite slow and
the isochrones of different ages are packed closely together.
Small uncertainties in $\log T_{\rm eff}, M_V$ or metallicity
give rise to large differences in the age of the MS star.
In the age-metallicity plane this results in an increased scatter
of the MS stars. 
\hfill\break
Edv93ea removed from their sample stars which lied too close
to the ZAMS with errors  
\mbox{log(age)\muspc$>$\muspc0.15}.
This is a subjective operation,
because it partly depends on the exact definition of the 
ZAMS. As a consequence some
MS might have been overlooked or too many taken out.
\par
The older ages obtained by Edv93ea for some of the solar metallicity
MS stars (see Figs.~2\,\&\,3) and the inclusion of these stars in the
definition of the AMR results in a larger scatter
in the definition of the disc AMR.
The location of the MS turn-off tabulated 
by Bert94ea is used by us to distinguish a MS
from a SGB star. In this way we classified more stars
than Edv93ea as a MS star. The larger variance 
of the MS stars in Tables~1\mbox{\muspc\&\muspc}2 appears to 
justify the current approach.
\hfill\break
As argued in Sect.~2.2.2 our fitting method is not 
substantially different from the one used by Edv93ea. 
The reason why they identified a different amount of
stars too close to the MS is related to either of
the following:
\begin{list}{$-$}{\topsep=0pt\parsep=0pt}
\item the exact location of the ZAMS adopted in the definition 
of $\Delta M_V$ for different metallicities,
\item a broader MS band due to convective overshoot.
\end{list}
To some extent a different choice of the isochrone with metallicity
Z from a specific [Fe/H] or [Me/H] affects not only the age,
but in our case also the distinction between MS or SGB star.

\subsection{SGB stars}
The situation changes if a star is on the SGB, where the evolution
is relatively fast and the uncertainty in the age is considerably
smaller.  
These stars potentially define a genuine disc AMR.
Figure~4a shows the AMR for the SGB stars.
It still has a large scatter in age and/or metallicity,
but there is definitely a small slope of \mbox{$\sim$0.04 dex/Gyr}
present. For comparison we show in 
Fig.~4b\mbox{\muspc\&\muspc}4c the AMR
for the same stars, but now with respectively
the visually assigned ages and the ages from Edv93ea.
Figure~4b has a similar slope, while
Fig.~4c shows an even steeper slope of
\mbox{$\sim$0.07 dex/Gyr}.
The latter slope is also 
obtained when the Lutz-Kelker 
correction is applied to the absolute magnitude of the stars
prior to the computation of the age or when the 
distances are derived from the {\it Hipparcos}\/ parallax, see Fig.~7a. 
However, one should be aware of the caveat that the slope might
be shallower. Because some old, metal-rich stars might be absent
in the sample, due to the selection criteria used by Edv93ea. 
\par
\mbox{Figures~3, 4a\to{c}}, 6 and 7a\muspc\&\muspc{b} 
indicate that 
there is no apparent AMR for stars with an age less than 10~Gyr.
We are basically dealing with a large metallicity
spread among the stars. Only, if we consider the older 
stars a slope appears. 
\par
A comparison of \mbox{Figs.~4a\muspc\&\muspc4b} further indicates that there
is no significant difference in the AMR slope between the automatically
determined ages and the visually improved ages of the SGB stars.
Preference is given to the objectively determined ages for 
the SGB stars in Fig.~4a, which have an uncertainty in their age of less 
than 12\%, while the estimated uncertainty in the metallicity
is smaller than 0.07~dex. 
The slope in the AMR is not expected to be an artifact.
The large spread in age and/or metallicity in Figs.~4a and 
7a\mbox{\muspc\&\muspc}7b is likely real. 
It could imply that the AMR in Fig.~6
is a superposition of a multitude of relations due to
in-fall or past mergers events. 
\par 
\begin{figure}
\centerline{\vbox{\psfig{file=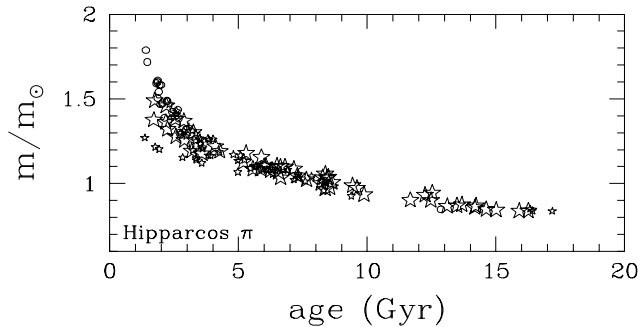,height=4.21cm,width=8.5cm}
}}
\caption{The age-mass relation for the MS and SGB stars in 
Tables~7\mbox{\muspc\&\muspc}8 with an
uncertainty in $\rm\overline{log(age)}\!<\!0.05$. 
The results were obtained with calculations
based on the {\it Hipparcos}\/ parallax.
We refer to Fig.~2 for the symbol description}
\label{mass}
\end{figure}

\subsection{Initial stellar masses}
Figure~\ref{mass} displays the ages with the corresponding masses obtained
for the reliable ($\sigma_\tau\!<\!0.05$)
MS stars from Table~7 and the SGB stars from Table~8. 
One expects a smooth and continuous 
relation, but between 10\to12~Gyr  
there is a hint of a slight discrepancy
in the distribution of the stars. It could be an indication
that stars with a larger age are younger and/or that the stars with
ages larger than \mbox{$\sim$\muspc8~Gyr} are possibly slightly older.
It might originate from the transition of isochrones with
convective overshoot to isochrones without it,
see Sect.~3.1 for the discussion about the
presence or absence of overshoot.
\hfill\break
On the other hand,
some stars with ages \mbox{$\ga$\muspc13~Gyr} could be 
systematically younger. Since the same procedure was
applied to determine the age, the origin should be due to
one or more input parameters, possibly a combination of [Fe/H] and
$T_{\rm eff}$. 
However, [Fe/H] is likely the dominant factor.  
Moreover, at \mbox{[Fe/H]\muspc$\la$\muspc\to0.4} 
Edv93ea found that
the stars are relatively over-abundant in the $\alpha$ elements.
In these cases we might have not used the optimum set of isochrones
to determine the age, which 
results in a small discrepancy in both the older ages
and the initial mass for these stars.

\subsection{In summary}
We have computed with the Bert94ea isochrones
new ages for the stellar sample defined by Edv93ea.
The revised values are systematically larger towards older ages.
The differences are considerably smaller when the 
distances are based on the {\it Hipparcos}\/ parallax.
The stars on the SGB define a disc AMR with a slope of 
\mbox{$\sim0.07$ dex/Gyr}. A comparable slope is obtained,
when the Edv93ea ages are adopted for the same stars.

\acknowledgements{A. Bressan, C. Chiosi and L. Girardi are acknowledged for
their comments and discussions. 
We further thank the referee for his stimulating suggestions and 
critical notes, which lead,
together with the comments made afterwards by Drs. B.~Gustafsson,
B.~Edvardsson and P.E.~Nissen
to a substantial improvement of
the presentation of the paper.
Bertelli acknowledges
the financial support received from the Italian Ministry of University,
Scientific Research and Technology (MURST)
and the National Council of Research.
Ng is supported by TMR grant ERBFMRX-CT96-0086 from the European Community.}

\vfill\typeset\eject
\input 6182_table3.tex
\addtocounter{page}{-1}
\vfill\eject
\input 6182_table4.tex
\input 6182_table5.tex
\input 6182_table6.tex
\input 6182_table7.tex
\input 6182_table8.tex

\end{document}

%% file: 6182_table3.tex
\begin{table*}[t]
\caption[]{The age of F- and G-type on or very near the 
main sequence stars in the solar neighbourhood. The age is
an average over the sensitivity analysis performed, see 
Sects.~2.2.3 \& 3.2 for additional details. 
Column~1 gives the identifier of the star, columns~2\&3
give the mean mass \& the standard deviation for this mean value,
columns~4\&5 give the average age together with its standard
deviation, and column~6 shows the number of good fits (out of
the maximum 7, see Sect.~2.2.3) over which the average is
obtained. In addition, $n\!=\!1$ indicates that
1 out of the 7 fits converged. The values in columns 2\&4
refer to the good fit, while the standard deviation is 
in that case calculated over the `good' (1) and
`second best' (6) estimates. A value $n\!=\!0$ indicates
not a good fit was obtained in all 7 possible cases
and the mean \& standard deviation refer in this case
to the `second best' value}
\begin{center}
\begin{tabular}{|lr|rrrr|c|c|lr|rrrr|c|}
\cline{1-7}\cline{9-15}
ID   & & m\ \ \null& $\sigma_m$\ \null& $\tau$\ \ \null& 
$\sigma_\tau$\ \null & $n$ & \hspace{3.8mm}&
ID   & & m\ \ \null& $\sigma_m$\ \null& $\tau$\ \ \null& 
$\sigma_\tau$\ \null & $n$ \\
\cline{1-7}\cline{9-15}
  HR &    140   & 1.31 & 0.04  & 9.293 & 0.115& 7  &&  
  HR &   6315   & 1.14 & 0.03  & 9.365 & 0.254& 6  \\
  HR &    219   & 0.97 & 0.03  & 8.957 & 1.337& 4  &&
  HR &   6409   & 1.52 & 0.06  & 9.265 & 0.032& 7  \\
  HR &    370   & 1.21 & 0.04  & 9.502 & 0.053& 7  &&
  HR &   6775   & 0.89 & 0.03  & 9.965 & 0.182& 7  \\
  HR &    458   & 1.28 & 0.04  & 9.433 & 0.038& 7  &&
  HR &   6907   & 1.35 & 0.04  & 9.298 & 0.063& 7  \\
  HR &    672   & 1.25 & 0.07  & 9.572 & 0.102& 6  &&
  HR &   7126   & 1.43 & 0.03  & 8.916 & 0.263& 7  \\ 
  HR &    784   & 1.20 & 0.02  & 9.182 & 0.258& 5  &&
  HR &   7560   & 1.28 & 0.05  & 9.473 & 0.035& 7  \\
  HR &    799   & 1.21 & 0.03  & 9.227 & 0.282& 6  &&
  HR &   7955   & 1.34 & 0.05  & 9.421 & 0.030& 7  \\
  HR &    962   & 1.32 & 0.05  & 9.471 & 0.037& 7  &&
  HR &   8181   & 0.90 & 0.03  & 9.893 & 0.231& 7  \\
  HR &   1010   & 0.96 & 0.03  & 9.730 & 0.505& 6  &&
  HR &   8472   & 1.54 & 0.06  & 9.312 & 0.040& 7  \\
  HR &   1173   & 1.44 & 0.04  & 9.062 & 0.177& 7  &&
  HR &   8885   & 1.39 & 0.04  & 9.255 & 0.062& 7  \\
  HR &   1257   & 1.42 & 0.07  & 9.386 & 0.067& 7  &&
  HD &   6434   & 0.85 & 0.03  & 9.945 & 0.245& 6  \\
  HR &   1489   & 1.22 & 0.06  & 9.592 & 0.088& 7  &&
  HD &  17548   & 0.88 & 0.01 & 10.048 & 0.072& 4  \\
  HR &   1687   & 1.43 & 0.04  & 8.923 & 0.367& 7  &&
  HD &  22879   & 0.79 & 0.02 & 10.089 & 0.115& 2  \\
  HR &   1780   & 1.15 & 0.03  & 9.304 & 0.318& 5  &&
  HD &  25704   & 0.80 & 0.03 & 10.038 & 0.169& 3  \\
  HR &   1983   & 1.22 & 0.03  & 9.219 & 0.232& 6  &&
  HD &  30649   & 0.82 & 0.03 & 10.246 & 0.020& 2  \\
  HR &   2047   & 1.06 & 0.02  & 9.531 & 0.234& 5  &&
  HD &  43947   & 0.96 & 0.03  & 9.942 & 0.080& 6  \\
  HR &   2220   & 1.34 & 0.03  & 9.098 & 0.157& 6  &&
  HD &  51929   & 0.83 & 0.03 & 10.073 & 0.063& 4  \\
  HR &   2493   & 0.98 & 0.02  & 9.874 & 0.092& 7  &&
  HD &  62301   & 0.84 & 0.02 & 10.091 & 0.025& 2  \\
  HR &   2721   & 0.99 & 0.04  & 8.987 & 1.741& 6  &&
  HD &  66573   & 0.80 & 0.03 & 10.136 & 0.182& 4  \\
  HR &   2943   & 1.51 & 0.05  & 9.235 & 0.027& 7  &&
  HD &  69611   & 0.81 & 0.03 & 10.237 & 0.022& 2  \\
  HR &   3018   & 0.82 & 0.03  & 8.733 & 0.277& 1  &&
  HD &  74011   & 0.82 & 0.03 & 10.238 & 0.844& 1  \\
  HR &   3578   & 0.79 & 0.03 & 10.156 & 0.150& 6  &&
  HD &  78747   & 0.80 & 0.04 & 10.120 & 0.250& 6  \\
  HR &   3954   & 1.40 & 0.07  & 9.366 & 0.076& 7  &&
  HD &  89707   & 0.94 & 0.04  & 9.750 & 0.497& 7  \\
  HR &   4012   & 1.38 & 0.05  & 9.419 & 0.030& 7  &&
  HD &  91347   & 0.86 & 0.03 & 10.057 & 0.157& 6  \\
  HR &   4067   & 1.40 & 0.05  & 9.326 & 0.028& 7  &&
  HD &  98553   & 0.91 & 0.04  & 9.840 & 0.271& 6  \\
  HR &   4395   & 1.51 & 0.06  & 9.273 & 0.029& 7  &&
  HD & 114762   & 0.78 & 0.02 & 10.214 & 0.000& 2  \\
  HR &   4529   & 1.37 & 0.05  & 9.434 & 0.032& 7  &&
  HD & 126512   & 0.83 & 0.03 & 10.229 & 0.841& 1  \\
  HR &   4533   & 1.36 & 0.04  & 9.189 & 0.131& 7  &&
  HD & 134169   & 0.82 & 0.04  & 8.006 & 0.004& 0  \\
  HR &   4540   & 1.31 & 0.05  & 9.429 & 0.032& 7  &&
  HD & 148211   & 0.83 & 0.04 & 10.192 & 0.052& 2  \\
  HR &   4657   & 0.93 & 0.03  & 9.723 & 0.415& 7  &&
  HD & 148816   & 0.78 & 0.03 & 10.184 & 0.824& 1  \\
  HR &   4688   & 1.37 & 0.04  & 9.289 & 0.050& 7  &&
  HD & 155358   & 0.81 & 0.04 & 10.230 & 0.038& 2  \\
  HR &   4767   & 1.08 & 0.03  & 9.599 & 0.187& 6  &&
  HD & 165401   & 0.85 & 0.02 & 10.083 & 0.074& 5  \\
  HR &   4785   & 0.98 & 0.04  & 9.774 & 0.244& 6  &&
  HD & 174912   & 0.85 & 0.04 & 10.050 & 0.220& 6  \\
  HR &   4845   & 0.81 & 0.02 & 10.189 & 0.095& 4  &&
  HD & 184499   & 0.86 & 0.03  & 9.475 & 0.556& 1  \\
  HR &   4903   & 1.44 & 0.04  & 9.420 & 0.035& 6  &&
  HD & 199289   & 0.78 & 0.03 & 10.042 & 0.140& 5  \\
  HR &   5011   & 1.24 & 0.05  & 9.551 & 0.042& 7  &&
  HD & 201891   & 0.79 & 0.01  & 9.904 & 0.077& 3  \\
  HR &   5235   & 1.53 & 0.04  & 9.337 & 0.036& 6  &&
  HD & 208906   & 0.85 & 0.04  & 9.958 & 0.217& 6  \\
  HR &   5323   & 1.36 & 0.05  & 9.411 & 0.030& 7  &&
  HD & 210752   & 0.82 & 0.03 & 10.137 & 0.132& 6  \\
  HR &   5542   & 1.30 & 0.08  & 9.534 & 0.091& 7  &&
  HD & 215257   & 0.86 & 0.02 & 10.036 & 0.063& 6  \\
  HR &   5698   & 1.44 & 0.05  & 9.363 & 0.030& 7  &&
  HD & 218504   & 0.86 & 0.04 & 10.135 & 0.031& 2  \\
  HR &   6243   & 1.66 & 0.07  & 9.219 & 0.042& 7  &&
     &          &      &       &       &      &    \\
\cline{1-7}\cline{9-15}
\end{tabular}
\end{center}
\end{table*}

%% file: 6182_table4.tex
\begin{table*}[t]
\caption[]{The age of F- and G-type 
sub-giant branch stars in the solar neighbourhood.
See caption of Table~3 for a description of columns~1\to6
and additional details.
In addition, column~7 refers to the visually assigned ages}
\begin{center}
\begin{tabular}{|lr|rrrr|cr|c|lr|rrrr|cr|}
\cline{1-8}\cline{10-17}
ID   & & m\ \ \null& $\sigma_m$\ \null& $\tau$\ \ \null& 
$\sigma_\tau$\ \null & $n$ &$\tau_v$\ \null & \hspace{3.8mm}&
ID   & & m\ \ \null& $\sigma_m$\ \null& $\tau$\ \ \null& 
$\sigma_\tau$\ \null & $n$ &$\tau_v$\ \null\\
\cline{1-8}\cline{10-17}
  HR &    17    & 1.06 & 0.04  & 9.790 & 0.086 & 5 & 9.75& &  
  HR &  5019    & 0.96 & 0.03 & 10.087 & 0.054 & 7 &10.09\\  
  HR &    33    & 1.09 & 0.03  & 9.774 & 0.057 & 6 & 9.68& &  
  HR &  5338    & 1.32 & 0.04  & 9.517 & 0.032 & 6 & 9.45\\  
  HR &    35    & 1.29 & 0.03  & 8.569 & 1.660 & 6 & 9.33& &  
  HR &  5353    & 1.34 & 0.01  & 9.545 & 0.009 & 5 & 9.54\\  
  HR &   107    & 1.22 & 0.06  & 9.550 & 0.106 & 7 & 9.50& &  
  HR &  5423    & 1.22 & 0.12  & 9.695 & 0.153 & 3 & 9.70\\  
  HR &   145    & 1.17 & 0.05  & 9.659 & 0.078 & 5 & 9.64& &  
  HR &  5447    & 1.25 & 0.03  & 9.349 & 0.086 & 7 & 9.38\\  
  HR &   203    & 1.00 & 0.04  & 9.975 & 0.062 & 7 & 9.98& &  
  HR &  5459    & 1.00 & 0.02  & 9.928 & 0.087 & 5 & 9.89\\  
  HR &   235    & 1.14 & 0.02  & 9.416 & 0.170 & 4 & 9.56& &  
  HR &  5691    & 1.27 & 0.08  & 9.535 & 0.096 & 7 & 9.50\\  
  HR &   244    & 1.31 & 0.05  & 9.497 & 0.069 & 7 & 9.45& &  
  HR &  5723    & 1.46 & 0.07  & 9.341 & 0.067 & 7 & 9.32\\  
  HR &   340    & 1.21 & 0.06  & 9.669 & 0.084 & 4 & 9.63& &  
  HR &  5868    & 1.10 & 0.03  & 9.815 & 0.048 & 4 & 9.73\\  
  HR &   366    & 1.24 & 0.07  & 9.522 & 0.115 & 7 & 9.47& &  
  HR &  5914    & 0.86 & 0.03 & 10.163 & 0.039 & 3 &10.20\\  
  HR &   368    & 1.39 & 0.07  & 9.390 & 0.069 & 7 & 9.34& &  
  HR &  5933    & 1.21 & 0.05  & 9.450 & 0.046 & 4 & 9.51\\  
  HR &   448    & 1.24 & 0.06  & 9.645 & 0.067 & 7 & 9.50& &  
  HR &  5968    & 0.93 & 0.03 & 10.083 & 0.042 & 6 &10.10\\  
  HR &   483    & 1.06 & 0.03  & 9.807 & 0.088 & 5 & 9.81& &  
  HR &  5996    & 1.26 & 0.06  & 9.603 & 0.078 & 6 & 9.49\\  
  HR &   573    & 1.08 & 0.03  & 9.717 & 0.079 & 6 & 9.72& &  
  HR &  6189    & 1.04 & 0.04  & 9.829 & 0.051 & 6 & 9.69\\  
  HR &   646    & 1.24 & 0.06  & 9.554 & 0.095 & 6 & 9.47& &  
  HR &  6202    & 1.38 & 0.07  & 9.406 & 0.061 & 5 & 9.34\\  
  HR &   720    & 1.01 & 0.04  & 9.958 & 0.051 & 7 & 9.96& &  
  HR &  6458    & 0.88 & 0.03 & 10.169 & 0.817 & 1 &10.26\\
  HR &   740    & 1.35 & 0.08  & 9.448 & 0.078 & 7 & 9.38& &  
  HR &  6541    & 1.29 & 0.07  & 9.529 & 0.082 & 7 & 9.43\\  
  HR &  1083    & 1.42 & 0.04  & 9.241 & 0.044 & 7 & 9.26& &  
  HR &  6569    & 1.31 & 0.03  & 9.367 & 0.050 & 7 & 9.36\\  
  HR &  1101    & 1.09 & 0.03  & 9.824 & 0.037 & 5 & 9.65& &  
  HR &  6598    & 0.97 & 0.04 & 10.019 & 0.069 & 7 &10.02\\  
  HR &  1294    & 0.98 & 0.04 & 10.029 & 0.057 & 7 &10.03& &  
  HR &  6649    & 1.02 & 0.03  & 9.877 & 0.069 & 7 & 9.88\\  
  HR &  1536    & 1.28 & 0.07  & 9.595 & 0.078 & 7 & 9.50& &  
  HR &  6701    & 1.31 & 0.05  & 9.512 & 0.047 & 4 & 9.43\\  
  HR &  1545    & 1.20 & 0.04  & 9.554 & 0.061 & 7 & 9.55& &  
  HR &  6850    & 1.27 & 0.04  & 9.415 & 0.033 & 7 & 9.43\\  
  HR &  1673    & 1.26 & 0.06  & 9.496 & 0.086 & 7 & 9.45& &  
  HR &  7061    & 1.37 & 0.10  & 9.449 & 0.102 & 7 & 9.36\\  
  HR &  1729    & 1.10 & 0.04  & 9.821 & 0.052 & 7 & 9.68& &  
  HR &  7232    & 0.99 & 0.03 & 10.027 & 0.049 & 7 &10.03\\  
  HR &  2141    & 0.98 & 0.02  & 9.976 & 0.042 & 7 & 9.98& &  
  HR &  7322    & 1.17 & 0.04  & 9.653 & 0.078 & 7 & 9.56\\  
  HR &  2233    & 1.25 & 0.04  & 9.557 & 0.063 & 7 & 9.46& &  
  HR &  7534    & 1.27 & 0.03  & 9.462 & 0.048 & 6 & 9.46\\  
  HR &  2354    & 1.13 & 0.05  & 9.780 & 0.088 & 7 & 9.67& &  
  HR &  7766    & 0.96 & 0.03  & 9.993 & 0.042 & 6 & 9.99\\  
  HR &  2530    & 1.22 & 0.06  & 9.513 & 0.108 & 6 & 9.48& &  
  HR &  7875    & 1.01 & 0.04  & 9.915 & 0.062 & 7 & 9.92\\  
  HR &  2548    & 1.31 & 0.05  & 9.444 & 0.067 & 7 & 9.42& &  
  HR &  8027    & 1.11 & 0.04  & 9.710 & 0.062 & 6 & 9.66\\  
  HR &  2601    & 1.00 & 0.04  & 9.924 & 0.064 & 7 & 9.92& &  
  HR &  8041    & 1.16 & 0.05  & 9.741 & 0.061 & 7 & 9.60\\  
  HR &  2835    & 0.96 & 0.03  & 9.901 & 0.079 & 7 & 9.93& &  
  HR &  8077    & 1.29 & 0.10  & 9.547 & 0.124 & 5 & 9.49\\  
  HR &  2883    & 0.85 & 0.02 & 10.146 & 0.041 & 6 &10.16& &  
  HR &  8354    & 1.05 & 0.04  & 9.800 & 0.056 & 7 & 9.66\\  
  HR &  2906    & 1.26 & 0.08  & 9.574 & 0.102 & 5 & 9.48& &  
  HR &  8665    & 1.10 & 0.01  & 9.757 & 0.031 & 5 & 9.66\\  
  HR &  3176    & 1.16 & 0.05  & 9.770 & 0.069 & 7 & 9.62& &  
  HR &  8697    & 1.21 & 0.06  & 9.613 & 0.093 & 6 & 9.49\\  
  HR &  3220    & 1.27 & 0.04  & 9.427 & 0.044 & 7 & 9.42& &  
  HR &  8729    & 1.10 & 0.04  & 9.847 & 0.060 & 7 & 9.69\\  
  HR &  3262    & 1.13 & 0.03  & 9.308 & 0.580 & 3 & 9.54& &  
  HR &  8805    & 1.36 & 0.05  & 9.329 & 0.025 & 7 & 9.34\\  
  HR &  3271    & 1.24 & 0.05  & 9.619 & 0.071 & 6 & 9.54& &  
  HR &  8853    & 1.21 & 0.07  & 9.590 & 0.111 & 7 & 9.56\\  
  HR &  3538    & 1.00 & 0.03  & 9.958 & 0.066 & 6 & 9.96& &  
  HR &  8969    & 1.18 & 0.06  & 9.614 & 0.103 & 6 & 9.57\\  
  HR &  3648    & 1.10 & 0.04  & 9.834 & 0.056 & 7 & 9.68& &  
  HD &  2615    & 1.11 & 0.04  & 9.730 & 0.062 & 7 & 9.58\\  
  HR &  3775    & 1.27 & 0.05  & 9.530 & 0.073 & 7 & 9.43& &  
  HD & 14938    & 1.07 & 0.03  & 9.809 & 0.042 & 4 & 9.73\\  
  HR &  3881    & 1.17 & 0.04  & 9.723 & 0.073 & 7 & 9.64& &  
  HD & 18768    & 0.85 & 0.03 & 10.191 & 0.049 & 6 &10.20\\  
  HR &  3951    & 1.19 & 0.05  & 9.718 & 0.074 & 7 & 9.55& &  
  HD & 38007    & 0.92 & 0.03 & 10.108 & 0.047 & 6 &10.12\\  
  HR &  4027    & 1.11 & 0.03  & 9.829 & 0.048 & 6 & 9.70& &  
  HD & 68284    & 0.94 & 0.04 & 10.016 & 0.069 & 7 &10.02\\  
  HR &  4039    & 1.01 & 0.02  & 9.808 & 0.105 & 6 & 9.81& &  
  HD & 78558    & 0.86 & 0.02 & 10.180 & 0.034 & 3 &10.20\\  
  HR &  4150    & 1.33 & 0.05  & 9.440 & 0.066 & 7 & 9.37& &  
  HD &130551    & 0.97 & 0.02  & 9.918 & 0.036 & 7 & 9.82\\  
  HR &  4158    & 1.08 & 0.03  & 9.793 & 0.073 & 6 & 9.71& &  
  HD &144172    & 1.12 & 0.04  & 9.723 & 0.048 & 7 & 9.61\\  
  HR &  4277    & 1.06 & 0.03  & 9.816 & 0.091 & 6 & 9.79& &  
  HD &157089    & 0.88 & 0.03  & 9.498 & 0.564 & 1 &10.25\\ 
  HR &  4285    & 1.06 & 0.04  & 9.869 & 0.067 & 7 & 9.87& &  
  HD &159307    & 1.07 & 0.04  & 9.775 & 0.067 & 7 & 9.78\\  
  HR &  4421    & 1.23 & 0.05  & 9.540 & 0.065 & 6 & 9.42& &  
  HD &188815    & 0.94 & 0.03  & 9.930 & 0.072 & 7 & 9.93\\  
  HR &  4683    & 1.10 & 0.05  & 9.718 & 0.075 & 7 & 9.61& &  
  HD &198044    & 1.03 & 0.04  & 9.865 & 0.064 & 6 & 9.82\\  
  HR &  4734    & 1.11 & 0.04  & 9.822 & 0.056 & 7 & 9.67& &  
  HD &200973    & 1.17 & 0.05  & 9.650 & 0.065 & 7 & 9.51\\  
  HR &  4981    & 1.38 & 0.09  & 9.426 & 0.081 & 7 & 9.37& &  
  HD &201099    & 0.90 & 0.02 & 10.100 & 0.035 & 6 &10.14\\  
  HR &  4983    & 1.11 & 0.03  & 9.562 & 0.175 & 5 & 9.63& &  
  HD &205294    & 1.13 & 0.03  & 9.726 & 0.043 & 5 & 9.58\\  
  HR &  4989    & 1.14 & 0.04  & 9.623 & 0.070 & 3 & 9.61& &  
  HD &221830    & 0.86 & 0.03 & 10.195 & 0.046 & 6 &10.20\\  
\cline{1-8}\cline{10-17}
\end{tabular}
\end{center}
\end{table*}

%% file: 6182_table5.tex
\begin{table*}[t]
\caption[]{Similar to Table 3, but now with the Lutz-Kelker
(1973) correction ($n$\muspc=\muspc4, $\sigma_\pi/\pi$\muspc=\muspc0.15;
Hanson 1979, equation 31) 
applied to the absolute magnitude of the 
F- and G-type stars on or near the main sequence
prior to the calculation of the age}
\begin{center}
\begin{tabular}{|lr|rrrr|c|c|lr|rrrr|c|}
\cline{1-7}\cline{9-15}
ID   & & m\ \ \null& $\sigma_m$\ \null& $\tau$\ \ \null& 
$\sigma_\tau$\ \null & $n$ & \hspace{3.8mm}&
ID   & & m\ \ \null& $\sigma_m$\ \null& $\tau$\ \ \null& 
$\sigma_\tau$\ \null & $n$ \\
\cline{1-7}\cline{9-15}
 HR &    140   & 1.36 & 0.05  & 9.344 & 0.030& 7  &&
 HR &   5542   & 1.36 & 0.08  & 9.485 & 0.076& 7  \\
 HR &    219   & 0.96 & 0.03  & 9.814 & 0.197& 6  &&
 HR &   5698   & 1.50 & 0.07  & 9.333 & 0.051& 7  \\
 HR &    370   & 1.27 & 0.05  & 9.495 & 0.037& 7  &&
 HR &   5723   & 1.55 & 0.10  & 9.288 & 0.086& 7  \\
 HR &    458   & 1.35 & 0.05  & 9.419 & 0.030& 7  &&
 HR &   6243   & 1.73 & 0.05  & 9.175 & 0.034& 6  \\
 HR &    672   & 1.34 & 0.06  & 9.487 & 0.051& 5  &&
 HR &   6409   & 1.60 & 0.06  & 9.223 & 0.037& 7  \\
 HR &    784   & 1.23 & 0.03  & 9.298 & 0.289& 7  &&
 HR &   6907   & 1.41 & 0.05  & 9.318 & 0.028& 7  \\
 HR &    962   & 1.39 & 0.05  & 9.433 & 0.033& 7  &&
 HR &   7061   & 1.51 & 0.06  & 9.327 & 0.037& 7  \\
 HR &   1010   & 0.96 & 0.03  & 9.952 & 0.091& 7  &&
 HR &   7126   & 1.48 & 0.04  & 9.074 & 0.130& 7  \\
 HR &   1173   & 1.50 & 0.05  & 9.158 & 0.046& 7  &&
 HR &   7560   & 1.34 & 0.05  & 9.445 & 0.034& 7  \\
 HR &   1257   & 1.50 & 0.06  & 9.334 & 0.051& 7  &&
 HR &   7955   & 1.40 & 0.05  & 9.396 & 0.030& 7  \\
 HR &   1489   & 1.28 & 0.07  & 9.542 & 0.088& 6  &&
 HR &   8181   & 0.90 & 0.01  & 9.990 & 0.051& 3  \\
 HR &   1687   & 1.48 & 0.05  & 9.115 & 0.073& 7  &&
 HR &   8472   & 1.61 & 0.05  & 9.268 & 0.036& 6  \\
 HR &   2220   & 1.38 & 0.04  & 9.215 & 0.087& 7  &&
 HR &   8885   & 1.45 & 0.05  & 9.274 & 0.027& 7  \\
 HR &   2721   & 0.98 & 0.03  & 9.776 & 0.304& 7  &&
 HD &   6434   & 0.86 & 0.02  &10.065 & 0.054& 4  \\
 HR &   2943   & 1.59 & 0.06  & 9.207 & 0.031& 7  &&
 HD &  22879   & 0.78 & 0.03  &10.166 & 0.818& 1  \\
 HR &   3018   & 0.80 & 0.04  & 8.754 & 0.285& 0  &&
 HD &  25704   & 0.76 & 0.01  &10.272 & 0.004& 2  \\
 HR &   3578   & 0.80 & 0.02  &10.176 & 0.020& 2  &&
 HD &  51929   & 0.84 & 0.02  &10.114 & 0.026& 2  \\
 HR &   3954   & 1.49 & 0.06  & 9.314 & 0.033& 7  &&
 HD &  62301   & 0.86 & 0.02  &10.155 & 0.039& 2  \\
 HR &   4012   & 1.46 & 0.06  & 9.379 & 0.035& 7  &&
 HD &  66573   & 0.79 & 0.02  &10.246 & 0.000& 2  \\
 HR &   4067   & 1.48 & 0.06  & 9.305 & 0.031& 7  &&
 HD &  78747   & 0.80 & 0.02  &10.192 & 0.000& 2  \\
 HR &   4395   & 1.59 & 0.06  & 9.230 & 0.036& 7  &&
 HD &  98553   & 0.90 & 0.03  &10.008 & 0.158& 7  \\
 HR &   4529   & 1.44 & 0.05  & 9.395 & 0.035& 7  &&
 HD & 114762   & 0.82 & 0.03  &10.203 & 0.831& 1  \\
 HR &   4533   & 1.42 & 0.05  & 9.258 & 0.042& 7  &&
 HD & 134169   & 0.82 & 0.03  &10.209 & 0.056& 2  \\
 HR &   4540   & 1.37 & 0.05  & 9.410 & 0.029& 7  &&
 HD & 148816   & 0.84 & 0.02  & 9.728 & 0.652& 1  \\
 HR &   4688   & 1.44 & 0.05  & 9.296 & 0.030& 7  &&
 HD & 174912   & 0.86 & 0.01  &10.123 & 0.044& 3  \\
 HR &   4785   & 0.97 & 0.03  & 9.916 & 0.136& 7  &&
 HD & 199289   & 0.82 & 0.02  & 9.866 & 0.039& 1  \\
 HR &   4845   & 0.81 & 0.03  &10.242 & 0.017& 2  &&
 HD & 201891   & 0.79 & 0.04  & 9.856 & 0.364& 3  \\
 HR &   4981   & 1.50 & 0.06  & 9.334 & 0.044& 7  &&
 HD & 208906   & 0.83 & 0.03  &10.082 & 0.140& 6  \\
 HR &   5011   & 1.29 & 0.07  & 9.522 & 0.075& 5  &&
 HD & 210752   & 0.84 & 0.02  &10.158 & 0.023& 3  \\
 HR &   5235   & 1.56 & 0.08  & 9.316 & 0.072& 3  &&
 HD & 215257   & 0.87 & 0.02  &10.061 & 0.011& 2  \\
 HR &   5323   & 1.43 & 0.05  & 9.379 & 0.031& 7  &&
    &          &      &       &       &      &    \\
\cline{1-7}\cline{9-15}
\end{tabular}
\end{center}
\end{table*}

%% file: 6182_table6.tex
\begin{table*}[t]
\caption[]{Similar to Table 5, but now for
F- and G-type stars on the sub-giant branch}
\begin{center}
\begin{tabular}{|lr|rrrr|c|c|lr|rrrr|c|}
\cline{1-7}\cline{9-15}
ID   & & m\ \ \null& $\sigma_m$\ \null& $\tau$\ \ \null& 
$\sigma_\tau$\ \null & $n$ & \hspace{3.8mm}&
ID   & & m\ \ \null& $\sigma_m$\ \null& $\tau$\ \ \null& 
$\sigma_\tau$\ \null & $n$ \\
\cline{1-7}\cline{9-15}
 HR &     17   & 1.10 & 0.04  & 9.749 & 0.061& 5  &&
 HR &    368   & 1.47 & 0.10  & 9.342 & 0.084& 7  \\
 HR &     33   & 1.14 & 0.06  & 9.724 & 0.072& 4  &&
 HR &    448   & 1.32 & 0.06  & 9.556 & 0.063& 7  \\
 HR &     35   & 1.33 & 0.04  & 9.322 & 0.039& 6  &&
 HR &    483   & 1.08 & 0.03  & 9.842 & 0.054& 7  \\
 HR &    107   & 1.26 & 0.06  & 9.516 & 0.093& 6  &&
 HR &    573   & 1.10 & 0.04  & 9.736 & 0.065& 6  \\
 HR &    145   & 1.18 & 0.06  & 9.677 & 0.092& 6  &&
 HR &    646   & 1.29 & 0.05  & 9.501 & 0.046& 4  \\
 HR &    203   & 1.05 & 0.04  & 9.906 & 0.057& 6  &&
 HR &    720   & 1.06 & 0.04  & 9.884 & 0.064& 7  \\
 HR &    235   & 1.15 & 0.02  & 9.577 & 0.116& 6  &&
 HR &    740   & 1.45 & 0.10  & 9.366 & 0.087& 7  \\
 HR &    244   & 1.35 & 0.06  & 9.491 & 0.060& 7  &&
 HR &    799   & 1.23 & 0.03  & 9.369 & 0.164& 7  \\
 HR &    340   & 1.37 & 0.07  & 9.468 & 0.054& 3  &&
 HR &   1083   & 1.48 & 0.05  & 9.251 & 0.022& 7  \\
 HR &    366   & 1.30 & 0.07  & 9.471 & 0.080& 7  &&
 HR &   1101   & 1.14 & 0.04  & 9.761 & 0.054& 4  \\
\cline{1-7}\cline{9-15}
\end{tabular}
\end{center}
\end{table*}

\vfill\eject
\addtocounter{table}{-1}
\begin{table*}[t]
\caption[]{\it Continued ...}
\begin{center}
\begin{tabular}{|lr|rrrr|c|c|lr|rrrr|c|}
\cline{1-7}\cline{9-15}
ID   & & m\ \ \null& $\sigma_m$\ \null& $\tau$\ \ \null& 
$\sigma_\tau$\ \null & $n$ & \hspace{3.8mm}&
ID   & & m\ \ \null& $\sigma_m$\ \null& $\tau$\ \ \null& 
$\sigma_\tau$\ \null & $n$ \\
\cline{1-7}\cline{9-15}
 HR &   1294   & 1.03 & 0.05  & 9.950 & 0.069& 6  &&
 HR &   6202   & 1.45 & 0.09  & 9.354 & 0.074& 5  \\
 HR &   1536   & 1.34 & 0.06  & 9.532 & 0.062& 6  &&
 HR &   6315   & 1.16 & 0.03  & 9.512 & 0.136& 7  \\
 HR &   1545   & 1.21 & 0.04  & 9.583 & 0.076& 6  &&
 HR &   6458   & 0.89 & 0.03  &10.162 & 0.047& 6  \\
 HR &   1673   & 1.31 & 0.08  & 9.471 & 0.091& 7  &&
 HR &   6541   & 1.33 & 0.05  & 9.491 & 0.049& 5  \\
 HR &   1729   & 1.14 & 0.03  & 9.784 & 0.046& 6  &&
 HR &   6569   & 1.38 & 0.04  & 9.347 & 0.045& 7  \\
 HR &   1780   & 1.18 & 0.03  & 9.504 & 0.127& 7  &&
 HR &   6598   & 1.03 & 0.05  & 9.918 & 0.080& 7  \\
 HR &   1983   & 1.25 & 0.03  & 9.357 & 0.141& 7  &&
 HR &   6649   & 1.04 & 0.04  & 9.864 & 0.046& 7  \\
 HR &   2047   & 1.08 & 0.03  & 9.421 & 0.662& 5  &&
 HR &   6701   & 1.35 & 0.05  & 9.482 & 0.051& 5  \\
 HR &   2141   & 1.01 & 0.03  & 9.944 & 0.040& 6  &&
 HR &   6775   & 0.90 & 0.01  &10.030 & 0.045& 2  \\
 HR &   2233   & 1.32 & 0.07  & 9.487 & 0.072& 7  &&
 HR &   6850   & 1.32 & 0.07  & 9.410 & 0.083& 7  \\
 HR &   2354   & 1.17 & 0.05  & 9.729 & 0.063& 7  &&
 HR &   7232   & 1.04 & 0.04  & 9.957 & 0.063& 7  \\
 HR &   2493   & 0.98 & 0.03  & 9.932 & 0.054& 5  &&
 HR &   7322   & 1.20 & 0.06  & 9.620 & 0.095& 6  \\
 HR &   2530   & 1.24 & 0.06  & 9.522 & 0.087& 6  &&
 HR &   7534   & 1.31 & 0.05  & 9.451 & 0.059& 7  \\
 HR &   2548   & 1.35 & 0.09  & 9.437 & 0.086& 7  &&
 HR &   7766   & 1.00 & 0.04  & 9.940 & 0.052& 7  \\
 HR &   2601   & 1.05 & 0.04  & 9.838 & 0.068& 7  &&
 HR &   7875   & 1.06 & 0.04  & 9.847 & 0.052& 6  \\
 HR &   2835   & 0.97 & 0.03  & 9.921 & 0.046& 6  &&
 HR &   8027   & 1.13 & 0.03  & 9.711 & 0.044& 7  \\
 HR &   2883   & 0.89 & 0.02  &10.078 & 0.045& 6  &&
 HR &   8041   & 1.23 & 0.05  & 9.656 & 0.067& 7  \\
 HR &   2906   & 1.32 & 0.06  & 9.520 & 0.059& 7  &&
 HR &   8077   & 1.32 & 0.09  & 9.529 & 0.098& 7  \\
 HR &   3176   & 1.23 & 0.06  & 9.673 & 0.075& 7  &&
 HR &   8354   & 1.11 & 0.05  & 9.720 & 0.069& 6  \\
 HR &   3220   & 1.32 & 0.05  & 9.428 & 0.067& 7  &&
 HR &   8665   & 1.16 & 0.05  & 9.693 & 0.080& 6  \\
 HR &   3262   & 1.15 & 0.03  & 9.548 & 0.120& 3  &&
 HR &   8697   & 1.26 & 0.05  & 9.566 & 0.065& 4  \\
 HR &   3271   & 1.31 & 0.06  & 9.557 & 0.064& 5  &&
 HR &   8729   & 1.17 & 0.05  & 9.757 & 0.071& 7  \\
 HR &   3538   & 1.03 & 0.03  & 9.946 & 0.044& 7  &&
 HR &   8805   & 1.42 & 0.07  & 9.340 & 0.069& 7  \\
 HR &   3648   & 1.15 & 0.03  & 9.772 & 0.048& 6  &&
 HR &   8853   & 1.24 & 0.06  & 9.585 & 0.092& 7  \\
 HR &   3775   & 1.33 & 0.10  & 9.483 & 0.092& 6  &&
 HR &   8969   & 1.23 & 0.04  & 9.583 & 0.073& 5  \\
 HR &   3881   & 1.19 & 0.04  & 9.706 & 0.055& 5  &&
 HD &   2615   & 1.17 & 0.05  & 9.640 & 0.073& 7  \\
 HR &   3951   & 1.26 & 0.05  & 9.631 & 0.064& 6  &&
 HD &  14938   & 1.10 & 0.04  & 9.771 & 0.052& 7  \\
 HR &   4027   & 1.16 & 0.06  & 9.758 & 0.079& 6  &&
 HD &  17548   & 0.87 & 0.02  &10.111 & 0.040& 7  \\
 HR &   4039   & 1.03 & 0.04  & 9.855 & 0.072& 6  &&
 HD &  18768   & 0.89 & 0.04  &10.115 & 0.074& 7  \\
 HR &   4150   & 1.42 & 0.08  & 9.376 & 0.071& 7  &&
 HD &  30649   & 0.86 & 0.02  &10.185 & 0.040& 2  \\
 HR &   4158   & 1.11 & 0.03  & 9.761 & 0.038& 7  &&
 HD &  38007   & 0.97 & 0.04  &10.035 & 0.074& 7  \\
 HR &   4277   & 1.08 & 0.04  & 9.833 & 0.082& 7  &&
 HD &  43947   & 0.99 & 0.04  & 9.957 & 0.068& 7  \\
 HR &   4285   & 1.11 & 0.04  & 9.796 & 0.056& 6  &&
 HD &  68284   & 1.00 & 0.05  & 9.917 & 0.078& 7  \\
 HR &   4421   & 1.29 & 0.06  & 9.472 & 0.074& 6  &&
 HD &  69611   & 0.86 & 0.02  &10.175 & 0.040& 2  \\
 HR &   4657   & 0.92 & 0.02  & 9.926 & 0.079& 7  &&
 HD &  74011   & 0.83 & 0.02  &10.224 & 0.046& 5  \\
 HR &   4683   & 1.14 & 0.05  & 9.676 & 0.059& 6  &&
 HD &  78558   & 0.90 & 0.02  &10.107 & 0.048& 6  \\
 HR &   4734   & 1.17 & 0.05  & 9.742 & 0.067& 7  &&
 HD &  89707   & 0.93 & 0.02  & 9.979 & 0.080& 6  \\
 HR &   4767   & 1.09 & 0.02  & 9.689 & 0.110& 6  &&
 HD &  91347   & 0.89 & 0.01  &10.083 & 0.060& 2  \\
 HR &   4903   & 1.46 & 0.06  & 9.416 & 0.068& 6  &&
 HD & 126512   & 0.84 & 0.02  &10.214 & 0.045& 5  \\
 HR &   4983   & 1.14 & 0.03  & 9.600 & 0.072& 6  &&
 HD & 130551   & 1.00 & 0.03  & 9.881 & 0.045& 7  \\
 HR &   4989   & 1.17 & 0.04  & 9.643 & 0.074& 7  &&
 HD & 144172   & 1.17 & 0.04  & 9.653 & 0.059& 7  \\
 HR &   5019   & 1.00 & 0.03  &10.028 & 0.048& 6  &&
 HD & 148211   & 0.86 & 0.02  &10.146 & 0.043& 6  \\
 HR &   5338   & 1.44 & 0.06  & 9.420 & 0.050& 5  &&
 HD & 155358   & 0.84 & 0.02  &10.184 & 0.042& 5  \\
 HR &   5353   & 1.43 & 0.12  & 9.468 & 0.102& 3  &&
 HD & 157089   & 0.87 & 0.03  &10.155 & 0.046& 6  \\
 HR &   5423   & 1.33 & 0.01  & 9.557 & 0.009& 4  &&
 HD & 159307   & 1.12 & 0.04  & 9.700 & 0.054& 6  \\
 HR &   5447   & 1.30 & 0.04  & 9.357 & 0.033& 7  &&
 HD & 165401   & 0.89 & 0.02  & 9.997 & 0.028& 2  \\
 HR &   5459   & 1.04 & 0.03  & 9.912 & 0.046& 5  &&
 HD & 184499   & 0.84 & 0.03  &10.203 & 0.049& 6  \\
 HR &   5691   & 1.31 & 0.06  & 9.524 & 0.065& 7  &&
 HD & 188815   & 0.96 & 0.02  & 9.938 & 0.037& 7  \\
 HR &   5868   & 1.13 & 0.04  & 9.780 & 0.056& 7  &&
 HD & 198044   & 1.05 & 0.03  & 9.849 & 0.044& 6  \\
 HR &   5914   & 0.89 & 0.03  &10.115 & 0.044& 6  &&
 HD & 200973   & 1.25 & 0.03  & 9.544 & 0.031& 5  \\
 HR &   5933   & 1.24 & 0.06  & 9.528 & 0.098& 6  &&
 HD & 201099   & 0.92 & 0.03  &10.066 & 0.050& 7  \\
 HR &   5968   & 0.97 & 0.04  &10.037 & 0.055& 7  &&
 HD & 205294   & 1.22 & 0.04  & 9.611 & 0.053& 6  \\
 HR &   5996   & 1.35 & 0.08  & 9.521 & 0.073& 6  &&
 HD & 218504   & 0.88 & 0.02  &10.105 & 0.043& 6  \\
 HR &   6189   & 1.11 & 0.05  & 9.744 & 0.065& 7  &&
 HD & 221830   & 0.90 & 0.04  &10.112 & 0.064& 7  \\
\cline{1-7}\cline{9-15}
\end{tabular}
\end{center}
\end{table*}

%% file: 6182_table7.tex
\begin{table*}[t]
\caption[]{The age of F- and G-type on or very near the 
main sequence stars in the solar neighbourhood. 
The age is
an average over the sensitivity analysis performed, see 
Sects.~2.2.3 \& 3.2 for additional details. 
The distances of the stars are based on the Hipparcos 
parallax (ESA 1997).
Column~1 gives the identifier of the star, columns~2\&3
give the mean mass \& the standard deviation for this mean value,
columns~4\&5 give the average age together with its standard
deviation, and column~6 shows the number of good fits (out of
the maximum 7, see Sect.~2.2.3) over which the average is
obtained. In addition, $n\!=\!1$ indicates that
1 out of the 7 fits converged. The values in columns 2\&4
refer to the good fit, while the standard deviation is 
in that case calculated over the `good' (1) and
`second best' (6) estimates. A value $n\!=\!0$ indicates
not a good fit was obtained in all 7 possible cases
and the mean \& standard deviation refer in this case
to the `second best' value}
\begin{center}
\begin{tabular}{|lr|rrrr|c|c|lr|rrrr|c|}
\cline{1-7}\cline{9-15}
ID   & & m\ \ \null& $\sigma_m$\ \null& $\tau$\ \ \null& 
$\sigma_\tau$\ \null & $n$ & \hspace{3.8mm}&
ID   & & m\ \ \null& $\sigma_m$\ \null& $\tau$\ \ \null& 
$\sigma_\tau$\ \null & $n$ \\
\cline{1-7}\cline{9-15}
 HR &    140   & 1.28 & 0.01  & 9.293 & 0.091& 7  &&
 HR &   5235   & 1.55 & 0.07  & 9.333 & 0.072& 5  \\
 HR &    219   & 0.95 & 0.02  & 9.911 & 0.094& 7  &&
 HR &   5323   & 1.42 & 0.01  & 9.386 & 0.017& 7  \\
 HR &    235   & 1.14 & 0.03  & 8.995 & 0.456& 6  &&
 HR &   5338   & 1.48 & 0.06  & 9.382 & 0.052& 5  \\
 HR &    458   & 1.29 & 0.01  & 9.438 & 0.034& 7  &&
 HR &   5542   & 1.30 & 0.06  & 9.529 & 0.085& 7  \\
 HR &    483   & 1.04 & 0.04  & 9.762 & 0.188& 3  &&
 HR &   5698   & 1.60 & 0.01  & 9.267 & 0.015& 7  \\
 HR &    672   & 1.22 & 0.04  & 9.584 & 0.082& 7  &&
 HR &   5723   & 1.54 & 0.02  & 9.280 & 0.015& 7  \\
 HR &    784   & 1.20 & 0.01  & 8.641 & 0.317& 3  &&
 HR &   6243   & 1.79 & 0.02  & 9.144 & 0.016& 7  \\
 HR &    962   & 1.32 & 0.01  & 9.471 & 0.024& 7  &&
 HR &   6315   & 1.14 & 0.02  & 9.448 & 0.138& 7  \\
 HR &   1010   & 0.98 & 0.03  & 9.544 & 0.499& 6  &&
 HR &   6409   & 1.72 & 0.02  & 9.163 & 0.015& 7  \\
 HR &   1173   & 1.45 & 0.01  & 9.128 & 0.052& 7  &&
 HR &   6458   & 0.86 & 0.01  & 8.006 & 0.002& 0  \\
 HR &   1257   & 1.49 & 0.03  & 9.346 & 0.034& 7  &&
 HR &   6907   & 1.39 & 0.01  & 9.326 & 0.030& 7  \\
 HR &   1294   & 0.94 & 0.02  &10.036 & 0.061& 2  &&
 HR &   7126   & 1.43 & 0.01  & 8.977 & 0.106& 7  \\
 HR &   1687   & 1.46 & 0.01  & 9.113 & 0.055& 7  &&
 HR &   7232   & 0.95 & 0.03  & 9.874 & 0.227& 7  \\
 HR &   1780   & 1.15 & 0.02  & 9.285 & 0.328& 7  &&
 HR &   7560   & 1.23 & 0.01  & 9.496 & 0.047& 7  \\
 HR &   1983   & 1.22 & 0.02  & 9.131 & 0.247& 7  &&
 HR &   7955   & 1.61 & 0.01  & 9.270 & 0.014& 7  \\
 HR &   2047   & 1.09 & 0.03  & 7.693 & 1.972& 3  &&
 HR &   8181   & 0.89 & 0.02  &10.004 & 0.069& 6  \\
 HR &   2220   & 1.32 & 0.02  & 7.794 & 2.711& 7  &&
 HR &   8472   & 1.58 & 0.01  & 9.286 & 0.014& 7  \\
 HR &   2493   & 0.97 & 0.02  & 9.890 & 0.061& 7  &&
 HR &   8885   & 1.50 & 0.02  & 9.268 & 0.021& 7  \\
 HR &   3018   & 0.83 & 0.00  & 8.004 & 0.001& 0  &&
 HD &   6434   & 0.86 & 0.01  &10.094 & 0.031& 4  \\
 HR &   3538   & 1.00 & 0.02  & 9.669 & 0.245& 6  &&
 HD &  17548   & 0.87 & 0.02  &10.078 & 0.055& 4  \\
 HR &   3578   & 0.85 & 0.02  & 9.101 & 0.414& 1  &&
 HD &  22879   & 0.79 & 0.00  & 9.584 & 0.599& 1  \\
 HR &   3954   & 1.59 & 0.02  & 9.255 & 0.017& 7  &&
 HD &  25704   & 0.83 & 0.01  & 8.004 & 0.003& 0  \\
 HR &   4012   & 1.58 & 0.02  & 9.299 & 0.020& 7  &&
 HD &  30649   & 0.85 & 0.01  & 8.005 & 0.002& 0  \\
 HR &   4039   & 1.02 & 0.04  & 9.830 & 0.087& 6  &&
 HD &  43947   & 0.96 & 0.03  & 9.959 & 0.073& 7  \\
 HR &   4067   & 1.47 & 0.01  & 9.310 & 0.021& 7  &&
 HD &  51929   & 0.86 & 0.01  & 8.886 & 0.335& 0  \\
 HR &   4395   & 1.26 & 0.01  & 8.005 & 0.002& 0  &&
 HD &  66573   & 0.79 & 0.02  &10.264 & 0.051& 3  \\
 HR &   4529   & 1.39 & 0.01  & 9.428 & 0.018& 7  &&
 HD &  78747   & 0.80 & 0.02  &10.239 & 0.049& 3  \\
 HR &   4533   & 1.33 & 0.01  & 9.140 & 0.104& 7  &&
 HD &  91347   & 0.86 & 0.02  &10.105 & 0.074& 5  \\
 HR &   4540   & 1.30 & 0.01  & 9.433 & 0.033& 7  &&
 HD &  98553   & 0.91 & 0.03  & 9.818 & 0.221& 7  \\
 HR &   4657   & 0.92 & 0.02  & 9.895 & 0.084& 6  &&
 HD & 114762   & 0.82 & 0.03  &10.202 & 0.829& 1  \\
 HR &   4688   & 1.47 & 0.02  & 9.293 & 0.026& 7  &&
 HD & 148816   & 0.82 & 0.00  &10.211 & 0.003& 2  \\
 HR &   4767   & 1.08 & 0.03  & 9.270 & 0.347& 6  &&
 HD & 165401   & 0.85 & 0.01  &10.109 & 0.045& 7  \\
 HR &   4785   & 0.97 & 0.03  & 9.855 & 0.128& 7  &&
 HD & 174912   & 0.83 & 0.03  &10.151 & 0.087& 7  \\
 HR &   4845   & 0.81 & 0.02  &10.233 & 0.059& 3  &&
 HD & 199289   & 0.79 & 0.01  & 8.003 & 0.002& 0  \\
 HR &   4903   & 1.44 & 0.05  & 9.422 & 0.048& 7  &&
 HD & 201891   & 0.79 & 0.01  & 8.003 & 0.003& 0  \\
 HR &   4981   & 1.48 & 0.05  & 9.348 & 0.046& 7  &&
 HD & 208906   & 0.82 & 0.02  &10.140 & 0.064& 6  \\
 HR &   4983   & 1.11 & 0.02  & 9.208 & 0.265& 6  &&
 HD & 210752   & 0.85 & 0.01  &10.113 & 0.797& 1  \\
 HR &   5019   & 0.92 & 0.03  & 9.955 & 0.151& 7  &&
 HD & 215257   & 0.88 & 0.01  &10.073 & 0.011& 2  \\
\cline{1-7}\cline{9-15}
\end{tabular}
\end{center}
\end{table*}

%% file: 6182_table8.tex
\begin{table*}[t]
\caption[]{The age of F- and G-type 
sub-giant branch stars in the solar neighbourhood.
See caption of Table~7 for additional details}
\begin{center}
\begin{tabular}{|lr|rrrr|c|c|lr|rrrr|c|}
\cline{1-7}\cline{9-15}
ID   & & m\ \ \null& $\sigma_m$\ \null& $\tau$\ \ \null& 
$\sigma_\tau$\ \null & $n$ & \hspace{3.8mm}&
ID   & & m\ \ \null& $\sigma_m$\ \null& $\tau$\ \ \null& 
$\sigma_\tau$\ \null & $n$ \\
\cline{1-7}\cline{9-15}
 HR &     17   & 1.08 & 0.03  & 9.785 & 0.056& 6  &&
 HR &   5691   & 1.26 & 0.05  & 9.581 & 0.064& 7  \\
 HR &     33   & 1.08 & 0.01  & 9.796 & 0.023& 6  &&
 HR &   5868   & 1.08 & 0.04  & 9.809 & 0.085& 6  \\
 HR &     35   & 1.27 & 0.02  & 9.133 & 0.174& 7  &&
 HR &   5914   & 0.97 & 0.01  & 9.982 & 0.013& 7  \\
 HR &    107   & 1.21 & 0.05  & 9.556 & 0.101& 7  &&
 HR &   5933   & 1.20 & 0.01  & 9.509 & 0.049& 5  \\
 HR &    145   & 1.10 & 0.01  & 9.778 & 0.018& 7  &&
 HR &   5968   & 0.93 & 0.02  &10.088 & 0.030& 7  \\
 HR &    203   & 1.03 & 0.02  & 9.926 & 0.022& 7  &&
 HR &   5996   & 1.10 & 0.04  & 9.755 & 0.103& 7  \\
 HR &    244   & 1.25 & 0.04  & 9.524 & 0.083& 7  &&
 HR &   6189   & 1.09 & 0.01  & 9.769 & 0.020& 5  \\
 HR &    340   & 1.75 & 0.03  & 9.122 & 0.026& 0  &&
 HR &   6202   & 1.38 & 0.06  & 9.408 & 0.048& 7  \\
 HR &    366   & 1.22 & 0.05  & 9.558 & 0.103& 7  &&
 HR &   6541   & 1.30 & 0.05  & 9.524 & 0.060& 6  \\
 HR &    368   & 1.46 & 0.04  & 9.341 & 0.038& 7  &&
 HR &   6569   & 1.32 & 0.01  & 9.351 & 0.022& 7  \\
 HR &    370   & 1.15 & 0.02  & 9.451 & 0.127& 7  &&
 HR &   6598   & 1.11 & 0.01  & 9.800 & 0.011& 4  \\
 HR &    448   & 1.18 & 0.01  & 9.724 & 0.020& 7  &&
 HR &   6649   & 1.05 & 0.02  & 9.863 & 0.020& 6  \\
 HR &    573   & 1.07 & 0.02  & 9.697 & 0.058& 6  &&
 HR &   6701   & 1.37 & 0.01  & 9.460 & 0.006& 7  \\
 HR &    646   & 1.26 & 0.04  & 9.548 & 0.068& 6  &&
 HR &   6775   & 0.90 & 0.02  &10.067 & 0.034& 7  \\
 HR &    720   & 1.10 & 0.02  & 9.833 & 0.021& 7  &&
 HR &   6850   & 1.28 & 0.01  & 9.409 & 0.025& 6  \\
 HR &    740   & 1.40 & 0.05  & 9.391 & 0.059& 6  &&
 HR &   7061   & 1.33 & 0.04  & 9.486 & 0.059& 7  \\
 HR &    799   & 1.20 & 0.02  & 9.283 & 0.174& 7  &&
 HR &   7322   & 1.27 & 0.05  & 9.544 & 0.067& 6  \\
 HR &   1083   & 1.38 & 0.01  & 9.232 & 0.043& 7  &&
 HR &   7534   & 1.26 & 0.01  & 9.463 & 0.033& 7  \\
 HR &   1101   & 1.10 & 0.01  & 9.812 & 0.021& 7  &&
 HR &   7766   & 1.01 & 0.02  & 9.937 & 0.026& 7  \\
 HR &   1489   & 1.12 & 0.02  & 9.553 & 0.159& 5  &&
 HR &   7875   & 1.10 & 0.01  & 9.783 & 0.014& 6  \\
 HR &   1536   & 1.16 & 0.05  & 9.706 & 0.101& 7  &&
 HR &   8027   & 1.10 & 0.03  & 9.753 & 0.053& 6  \\
 HR &   1545   & 1.18 & 0.02  & 9.554 & 0.034& 6  &&
 HR &   8041   & 1.06 & 0.03  & 9.864 & 0.080& 7  \\
 HR &   1673   & 1.24 & 0.06  & 9.545 & 0.106& 7  &&
 HR &   8077   & 1.26 & 0.04  & 9.586 & 0.050& 6  \\
 HR &   1729   & 1.05 & 0.03  & 9.864 & 0.065& 3  &&
 HR &   8354   & 1.03 & 0.02  & 9.822 & 0.020& 7  \\
 HR &   2141   & 0.99 & 0.02  & 9.975 & 0.033& 6  &&
 HR &   8665   & 1.17 & 0.01  & 9.681 & 0.016& 3  \\
 HR &   2233   & 1.30 & 0.04  & 9.513 & 0.048& 6  &&
 HR &   8697   & 1.22 & 0.01  & 9.613 & 0.017& 6  \\
 HR &   2354   & 1.08 & 0.03  & 9.840 & 0.063& 7  &&
 HR &   8729   & 1.01 & 0.01  & 9.880 & 0.036& 4  \\
 HR &   2530   & 1.19 & 0.02  & 9.503 & 0.034& 6  &&
 HR &   8805   & 1.36 & 0.01  & 9.337 & 0.024& 7  \\
 HR &   2548   & 1.30 & 0.03  & 9.439 & 0.070& 7  &&
 HR &   8853   & 1.15 & 0.02  & 9.526 & 0.130& 5  \\
 HR &   2601   & 1.19 & 0.02  & 9.630 & 0.022& 6  &&
 HR &   8969   & 1.19 & 0.05  & 9.629 & 0.104& 7  \\
 HR &   2721   & 1.47 & 0.02  & 9.367 & 0.012& 4  &&
 HD &   2615   & 1.08 & 0.02  & 9.767 & 0.031& 7  \\
 HR &   2835   & 0.96 & 0.03  & 9.920 & 0.065& 7  &&
 HD &  14938   & 1.07 & 0.02  & 9.816 & 0.028& 6  \\
 HR &   2883   & 0.94 & 0.01  & 9.994 & 0.019& 7  &&
 HD &  18768   & 1.00 & 0.02  & 9.914 & 0.026& 7  \\
 HR &   2906   & 1.43 & 0.05  & 9.414 & 0.037& 5  &&
 HD &  38007   & 1.04 & 0.02  & 9.917 & 0.028& 6  \\
 HR &   2943   & 1.49 & 0.01  & 9.242 & 0.030& 6  &&
 HD &  62301   & 0.85 & 0.01  &10.165 & 0.023& 7  \\
 HR &   3176   & 1.16 & 0.01  & 9.770 & 0.020& 7  &&
 HD &  68284   & 1.02 & 0.03  & 9.883 & 0.048& 7  \\
 HR &   3220   & 1.28 & 0.06  & 9.473 & 0.104& 7  &&
 HD &  69611   & 0.84 & 0.01  &10.215 & 0.025& 4  \\
 HR &   3262   & 1.14 & 0.02  & 9.526 & 0.094& 6  &&
 HD &  74011   & 0.84 & 0.02  &10.202 & 0.025& 7  \\
 HR &   3271   & 1.26 & 0.05  & 9.603 & 0.068& 5  &&
 HD &  78558   & 0.85 & 0.01  &10.213 & 0.018& 3  \\
 HR &   3648   & 1.09 & 0.01  & 9.854 & 0.021& 7  &&
 HD &  89707   & 0.93 & 0.02  & 9.972 & 0.078& 5  \\
 HR &   3775   & 1.41 & 0.05  & 9.414 & 0.061& 6  &&
 HD & 126512   & 0.88 & 0.02  &10.136 & 0.023& 7  \\
 HR &   3881   & 1.10 & 0.01  & 9.821 & 0.027& 7  &&
 HD & 130551   & 0.97 & 0.02  & 9.923 & 0.029& 7  \\
 HR &   3951   & 1.02 & 0.02  & 9.854 & 0.083& 6  &&
 HD & 134169   & 0.87 & 0.02  &10.117 & 0.034& 7  \\
 HR &   4027   & 1.04 & 0.01  & 9.929 & 0.027& 7  &&
 HD & 144172   & 1.13 & 0.02  & 9.715 & 0.021& 7  \\
 HR &   4150   & 1.31 & 0.05  & 9.459 & 0.083& 7  &&
 HD & 148211   & 0.86 & 0.02  &10.152 & 0.028& 7  \\
 HR &   4158   & 1.08 & 0.04  & 9.785 & 0.087& 7  &&
 HD & 155358   & 0.84 & 0.01  &10.176 & 0.024& 7  \\
 HR &   4277   & 1.05 & 0.02  & 9.801 & 0.076& 5  &&
 HD & 157089   & 0.87 & 0.01  &10.153 & 0.023& 7  \\
 HR &   4285   & 1.31 & 0.02  & 9.506 & 0.013& 3  &&
 HD & 159307   & 1.08 & 0.03  & 9.745 & 0.038& 6  \\
 HR &   4421   & 1.17 & 0.04  & 9.583 & 0.085& 7  &&
 HD & 184499   & 0.84 & 0.01  &10.211 & 0.019& 7  \\
 HR &   4683   & 1.09 & 0.04  & 9.737 & 0.063& 7  &&
 HD & 188815   & 0.98 & 0.02  & 9.932 & 0.024& 6  \\
 HR &   4734   & 1.04 & 0.03  & 9.894 & 0.077& 7  &&
 HD & 198044   & 1.03 & 0.03  & 9.865 & 0.067& 7  \\
 HR &   4989   & 1.15 & 0.00  & 9.622 & 0.016& 2  &&
 HD & 200973   & 1.14 & 0.04  & 9.696 & 0.054& 6  \\
 HR &   5011   & 1.16 & 0.01  & 9.593 & 0.053& 7  &&
 HD & 201099   & 0.90 & 0.01  &10.096 & 0.023& 7  \\
 HR &   5353   & 1.06 & 0.01  & 9.922 & 0.024& 7  &&
 HD & 205294   & 1.14 & 0.02  & 9.718 & 0.024& 5  \\
 HR &   5423   & 0.94 & 0.01  &10.098 & 0.041& 7  &&
 HD & 218504   & 0.87 & 0.02  &10.130 & 0.031& 6  \\
 HR &   5447   & 1.22 & 0.02  & 9.246 & 0.119& 6  &&
 HD & 221830   & 0.84 & 0.01  &10.235 & 0.024& 5  \\
 HR &   5459   & 1.01 & 0.03  & 9.914 & 0.079& 7  &&
    &          &      &       &       &      &    \\
\cline{1-7}\cline{9-15}
\end{tabular}
\end{center}
\end{table*}

%% file: 6182.bbl
\begin{thebibliography}{}
\bibitem [1992] {}
Bertelli G., Bressan A., Chiosi C., Fagotto F., Nasi E., 1994,
A\&AS 106, 275 (Bert94ea)
\bibitem [1992] {}
Chiosi C., Bertelli G., Bressan A., 1992, ARA\&A 30, 235
\bibitem [1993] {}
Edvardsson B., Andersen J., Gustafsson B., et~al., 1993,
A\&A 275, 101 (Edv93ea)
\bibitem [1997] {ESA97}
ESA, 1997, The {\it Hipparcos}\/ and {\it Tycho}\/ Catalogue, ESA SP-1200
\bibitem [1991] {}
Grevesse N., 1991, A\&A 242, 488
\bibitem [1976] {GO76}
Gr{\o}nbech B., Olsen E.H., 1976, A\&AS 25, 213
\bibitem [1979] {Hanson79}
Hanson R.B., 1979, MNRAS 186, 875
\bibitem [1991] {HW91}
Hoffleit D., Warren Jr. W.H., 1991, Bright Star Catalogue
(5$^{th}$ revised edition)
\bibitem [1977] {LAOL77}
Huebner W.F., Merts A.L., Magee N.H., Argo M.F., 1977,
Los Alamos Scientific Laboratory Report LA-6760-M
\bibitem [1992] {OPAL92a}
Iglesias C.A., Rodgers F.J., Wilson B.G., 1992, ApJ 397, 717
\bibitem [] {LK73}
Lutz T.E., Kelker D.H., 1973, PASP 85, 573
\bibitem [1983] {Olsen83}
Olsen, E.H., 1983, A\&AS 54, 55
\bibitem [1993] {Olsen93}
Olsen, E.H., 1993, A\&AS 102, 89
\bibitem [1989] {}
Pagel B.E.J., 1989, in `Evolutionary phenomena in Galaxies',
J. Beckman and B.E.J. Pagel (eds.), Cambridge University Press, 201  
\bibitem [1987] {POC87}
Perry C.L., Olsen, E.H., Crawford D.L., 1987, PASP 99, 1184
\bibitem [1986] {Press86}
Press W.H., et~al., 1986, {\it Numerical Recipes}
\bibitem [1992] {OPAL92b}
Rodgers F.J., Iglesias C.A., 1992, ApJS 79, 507
\bibitem [1988] {SN88}
Schuster W.J., Nissen P.E., 1988, A\&AS 73, 225
\bibitem [1953] {}
Trumpler R.J., Weaver H.F., 1953, {\it Statistical Astronomy},
University of California Press (Berkeley), 368
\bibitem [1983] {}
Vandenberg D.A., 1983, ApJS 51, 29
\bibitem [1985] {}
Vandenberg D.A., 1985, ApJS 58, 711 (Vdb85)
\end{thebibliography}
